\documentclass[preprint,10pt]{elsarticle}

\usepackage{lineno,hyperref}
\usepackage{amsmath}
\usepackage{subfigure}
\usepackage{caption}
\usepackage{xcolor}
\modulolinenumbers[5]

\journal{Journal of COMPUT METHOD APPL M}

\begin{document}

\begin{frontmatter}

\title{A Unified Finite Element Method \\
	for Fluid-Structure Interaction} 


\author[]{Yongxing Wang\corref{mycorrespondingauthor}}
\cortext[mycorrespondingauthor]{Corresponding author}
\ead{scywa@leeds.ac.uk}

\author[]{Peter Jimack}
\author[]{Mark Walkley}

\address{School of Computing, University of Leeds, Leeds, UK, LS2 9JT}

\begin{abstract}
In this article, we present a new unified finite element method (UFEM) for simulation of general Fluid-Structure interaction (FSI) which has the same generality and robustness as monolithic methods but is significantly more computationally efficient and easier to implement. Our proposed approach has similarities with classical immersed finite element methods (IFEMs), by approximating a single velocity and pressure field in the entire domain (i.e. occupied by fluid and solid) on a single mesh, but differs by treating the corrections due to the solid deformation on the left-hand side of the modified fluid flow equations (i.e. implicitly). The method is described in detail, followed by the presentation of multiple computational examples in order to validate it across a wide range of fluid and solid parameters and interactions.
\end{abstract}


\end{frontmatter}

\linenumbers

\section{Introduction }

Numerical simulation of fluid-structure interaction is a computational challenge because of its strong nonlinearity, especially when large deformation is considered. Based on how to couple the interaction between fluid and solid, existing numerical methods can be broadly categorized into two approaches: partitioned/segregated methods and monolithic/fully-coupled methods. Similarly, based on how to handle the mesh, they can also be broadly categorized into two further approaches: fitted mesh/conforming methods and unfitted/non-conforming mesh methods \cite{hou2012numerical}.

A fitted mesh means that the fluid and solid meshes match each other at the interface, and the nodes on the interface are shared by both the fluid and the solid, which leads to the fact that each interface node has both a fluid velocity and a solid velocity (or displacement) defined on it. It is apparent that the two velocities on each interface node should be consistent. There are typically two methods to handle this: partitioned/segregated methods \cite{kuttler2008fixed,Degroote_2009} and monolithic/fully-coupled methods \cite{Heil_2004,Heil_2008,Muddle_2012}. The former solve the fluid and solid equations sequentially and iterate until the velocities become consistent at the interface. These are more straightforward to implement but can lack robustness and may fail to converge when there is a significant energy exchange between the fluid and solid \cite{Degroote_2009}. The latter solve the fluid and solid equations simultaneously and often use a Lagrange Multiplier to weakly enforce the continuity of velocity on the interface \cite{Muddle_2012}. This has the advantage of achieving accurate and stable solutions, however the key computational challenge is to efficiently solve the large systems of nonlinear algebraic equations arising from the fully-coupled implicit discretization of the fluid and solid equations. Fitted mesh methods can accurately model wide classes of FSI problems, however maintaining the quality of the mesh for large solid deformations usually requires a combination of arbitrary Lagrangian-Eulerian (ALE) mesh movement and partial or full remeshing \cite{peterson1999solution}. These add to the computational expense and, when remeshing occurs, can lead to loss of conservation properties of the underlying discretization \cite{Walkley_2005}.

Unfitted mesh methods use two meshes to represent the fluid and solid separately and these do not generally conform to each other on the interface. In this case, the definition of the fluid problem may be extended to an augmented domain which includes the solid domain. Similarly to the fitted case, there are also two broad approaches to treat the solid domain: partitioned methods and monolithic methods. On an unfitted mesh, there is no clear boundary for the solid problem, so it is not easy to enforce the boundary condition and solve the solid equation. A wide variety of schemes have been proposed to address this issue, including the Immersed Finite Element Method (IFEM) \cite{zhang2004immersed,Zhang_2007,Wang_2009,Wang_2011,Wang_2013}, the Fictitious Domain (FD) method \cite{Glowinski_2001,Yu_2005,baaijens2001fictitious,Kadapa_2016}, and the mortar approach \cite{baaijens2001fictitious,Hesch_2014}. The IFEM developed from the Immersed Boundary method first introduced by Peskin \cite{peskin2002immersed}, and has had great success with applications in bioscience and biomedical fields. The classical IFEM does not solve solid equations at all. Instead, the solid equations are arranged on the right-hand side of the fluid equations as an FSI force, and these modified fluid equations are solved on the augmented domain (occupied by fluid and solid). There is also the Modified IFEM \cite{Wang_2013}, which solves the solid equations explicitly and iterates until convergence. Reference \cite{Glowinski_2001} presents a fractional scheme for a rigid body interacting with the fluid, whilst \cite{Yu_2005} introduces a fractional step scheme using Distributed Lagrange Multiplier (DLM)/FD for fluid/flexible-body interactions. In the case of monolithic methods, \cite{baaijens2001fictitious} uses a FD/mortar approach to couple the fluid and structure, but the coupling is limited to a line (2D) representing the structure. Reference \cite{Hesch_2014} uses a mortar approach to solve fluid interactions with deformable and rigid bodies, and \cite{Kadapa_2016} also solves a fully-coupled FSI system with hierarchical B-Spline grids. There are also other monolithic methods based on unfitted meshes \cite{Robinson_Mosher_2011,Hachem_2013}.

It can be seen that the major methods based on unfitted meshes either avoid solving the solid equations (IFEM) or solve them with additional variables (two velocity fields and Lagrange multiplier) in the solid domain. However, physically, there is only one velocity field in the solid domain. In this article, we develop a semi-explicit Unified FEM (UFEM) approach which only solves one velocity variable in the whole/augmented domain. We shall use unfitted meshes to introduce our UFEM, although the methodology can also be applied to fitted meshes.

The word ``unified" here has two meanings: (1) the equations for fluid and solid are unified in one equation in which only one velocity variable is solved; (2) a range of solid materials, from the very soft to the very hard, may be considered in this one scheme.

The term ``semi-explicit" also has two components: (1) we linearize the solid constitutive model (an incompressible neo-Hookean model) explicitly using the value from the last time step; (2) we couple the FSI interaction implicitly by arranging the solid information on the left-hand side of control equations.

The main idea of UFEM is as follows. We first discretize the control equations in time, re-write the solid equation in the form of a fluid equation (using the velocity as a variable rather than the displacement) and re-write the solid constitutive equation in the updated coordinate system. We then combine the fluid and solid equations and discretize them in an augmented domain. Finally the multi-physics problem is solved as a single field. 

The UFEM differs from the classical IFEM approach which puts all the solid model information from the last time step explicitly on the right-hand side of the fluid equations. This typically requires the use of a very small time step to simulate the whole FSI system. This IFEM approach works satisfactorily when the solid behaves like a fluid, such as a very soft solid, but can lead to significant errors when the solid behaves quite differently from the fluid, such as a hard solid. The UFEM scheme includes the solid information on the left-hand side and, as we will demonstrate, can simulate a wide range from very soft to very hard solids both accurately and efficiently.

As noted above, monolithic methods strongly couple the fluid and solid models, and discretize them into one implicit nonlinear equation system at each time step. The unknowns include velocity, displacement and a Lagrangian multiplier to enforce consistency of velocity on an interface (fitted mesh) \cite{Heil_2004,Heil_2008,Muddle_2012} or in a solid domain (unfitted mesh) \cite{baaijens2001fictitious,Kadapa_2016,Hesch_2014}. One may gain both a stable and an accurate solution from such fully-coupled schemes. However, it is clear that this strategy is very costly, especially for the unfitted mesh case, in which the so called mortar integrals are involved \cite{Hesch_2014}. The UFEM only solves for velocity as unknowns, which is cheaper, but does not lose stability or accuracy as shown by the numerical experiments reported in this paper.

The following sections are organized as follows. In section 2, the control equations and boundary conditions for fluid-structure interactions are introduced; In section 3, the weak form of the FSI system is presented based on the augmented fluid domain. In section 4, details of the linearization of the FSI equations are discussed and the numerical scheme is presented. In section 5, numerical examples are described to validate the proposed UFEM.

\section{Governing equations for FSI}
In the following context, let
\begin{equation}
\left(u,v\right)_\Omega=\int_{\Omega}uvd\Omega,
\end{equation}
where $u$ and $v$ are functions defined in domain $\Omega$.

All subscripts, such as $i$, $j$, and $k$, represent spatial dimension. If they are repeated in one term (including the bracket defined in (1)), it implies summation over the spatial dimension; if they are not repeated, they take the value 1 and 2 for 2D, and 1 to 3 for 3D. All superscripts are used to distinguish fluid and solid ($f$ and $s$ respectively), distinguish different boundaries ($\Gamma^D$ and $\Gamma^N$) or represent time step $\left(n\right)$. For example, $u_i^f$ and $u_i^s$ denote the velocity components of fluid and solid respectively, $\sigma_{ij}^f$ and $\sigma_{ij}^s$ denote the stress tensor components of fluid and solid respectively, and $\left(u_i^s\right)^n$ is a solid velocity component at time $t^n$.

\begin{figure}[h!]
\centering
\includegraphics[width=3in,angle=0]{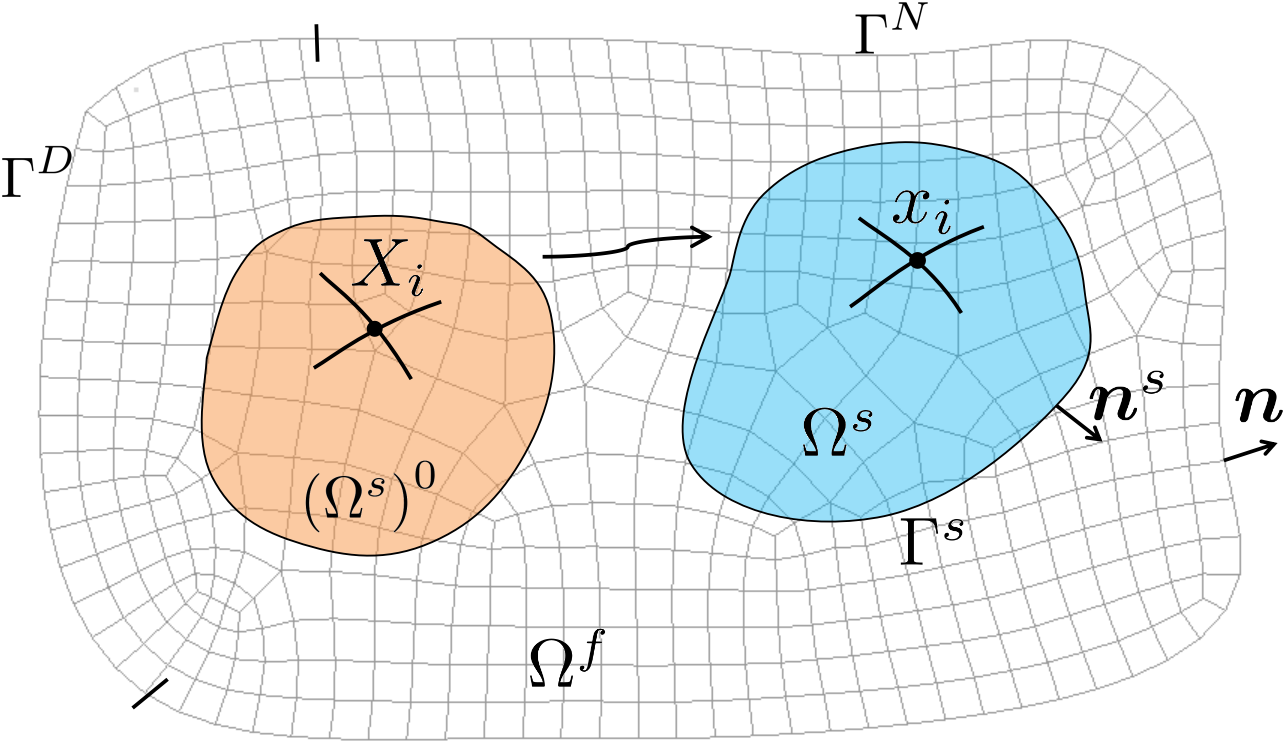}
\captionsetup{justification=centering}
\caption {\scriptsize Schematic diagram of FSI, $\Omega=\Omega^f\cup \Omega^s$, $\Gamma=\Gamma^D\cup \Gamma^N$.} 
\label{fig1}
\end{figure}

In our model we assume an incompressible fluid governed by the following equations in $\Omega^f$ as shown in Figure \ref{fig1}:
\begin{equation} \label{fluid1}
\rho^f\frac{Du_i^f}{Dt}
-\frac{\partial \sigma_{ij}^f}{\partial x_j}=\rho^f g_i,
\end{equation}

\begin{equation} \label{fluid2}
\frac{\partial u_j^f}{\partial x_j}=0,
\end{equation}

\begin{equation} \label{fluid3}
\sigma_{ij}^f=\mu^f \left(\frac{\partial u_i^f}{\partial x_j}+\frac{\partial u_j^f}{\partial x_i} \right)-p^f\delta_{ij}
=\tau_{ij}^f-p^f\delta_{ij}.
\end{equation}

We also assume an incompressible solid that is governed by the following equations in $\Omega^s$ as shown in Figure \ref{fig1}:
\begin{equation} \label{solid1}
\rho^s \frac{Du_i^s}{Dt} -\frac{\partial \sigma_{ij}^s}{\partial x_j}=\rho^s g_i,
\end{equation}

\begin{equation} \label{solid2}
det \left( {\bf F} \right)=1,
\end{equation}

\begin{equation} \label{solid3}
\sigma_{ij}^s=\mu^s \left(\frac{\partial x_i^s}{\partial X_k}\frac{\partial x_j^s}{\partial X_k} - \delta_{ij} \right)-p^s\delta_{ij}
=\tau_{ij}^s-p^s\delta_{ij}.
\end{equation}

In the above $\tau_{ij}^f$ and $\tau_{ij}^s$ are the deviatoric stress of the fluid and solid respectively, $\rho^f$ and $\rho^s$ are the density of the fluid and solid respectively, $\mu^f$ is the fluid viscosity, and $g_i$ is the acceleration due to gravity. Note that (\ref{solid1})-(\ref{solid3}) describe an incompressible neo-Hookean model that is based on \cite{baaijens2001fictitious} and is suitable for large displacements. In this model, $\mu^s$ is the shear modulus and $p^s$ is the pressure of the solid ($p^f$ being the fluid pressure in (\ref{fluid3})). We denote by $x_i$ the current coordinates of the solid or fluid, and by $X_i$ the reference coordinates of the solid, whilst ${\bf F}=\left[\frac{\partial x_i}{\partial X_j}\right]$ is the deformation tensor of the solid and $\frac{D}{Dt}$ represents the total derivative of time.

On the interface boundary $\Gamma^s$: 
 \begin{equation}\label{interfaceBC1}
 u_i^f=u_i^s,
 \end{equation}
 \begin{equation}\label{interfaceBC2}
 \sigma_{ij}^fn_j^s=\sigma_{ij}^sn_j^s,
 \end{equation}
where $n_j^s$ denotes the component of outward pointing unit normal, see Figure \ref*{fig1}.

Dirichlet and Neumann boundary conditions may be imposed for the fluid:

\begin{equation}
u_i^f=\bar{u}_i \quad on \quad  \Gamma^D,
\end{equation}

\begin{equation}\label{NeummanBC}
\sigma_{ij}^fn_j=\bar{h}_i \quad on \quad  \Gamma^N.
\end{equation}

Finally, initial conditions are typically set as:
\begin{equation} \label{initialcd}
 \left. u_i^f\right|_{t=0}=\left. u_i^s\right|_{t=0}=0,
\end{equation}
though they may differ from (\ref {initialcd}).

{\bf Remark 1} Using Jacobi's formula \cite{BODEWIG_2014}:
\begin{equation}
\frac{d}{dt}det\left({\bf F}\right)=det\left({\bf F}\right)tr\left({\bf F}^{-1}\frac{d {\bf F}}{dt}\right),
\end{equation}
we have
\begin{equation}
\frac{d}{dt}det\left({\bf F}\right)=det\left({\bf F}\right) \frac{\partial u_j^s}{\partial x_j},
\end{equation}
which, using (\ref{solid2}), gives
$$
\frac{\partial u_j^s}{\partial x_j}=0. \eqno \left(6^\prime \right)
$$
We choose that the reference configuration is the same as the initial configuration, so $(6^\prime)$ also implies (6). In our UFEM model, the incompressibility constraint $(6^\prime)$ will be used instead of (6).

\section{Weak form of FSI equations}
In order to obtain a weak formulation we define a combined trial space for velocity as:
$$
W=\left\{\left(u_i^f,u_i^s\right):u_i^f \in H^1\left(\Omega^f\right), u_i^s \in H^1\left(\Omega^s\right),
\left.u_i^s\right|_{\Gamma^s}=\left.u_i^f\right|_{\Gamma^s},
\left.u_i^f\right|_{\Gamma^D}=\bar{u}_i  \right\},
$$
with a corresponding combined test space for the velocity as:
$$
W_0=\left\{\left(v_i^f,v_i^s\right):v_i^f \in H^1\left(\Omega^f\right), v_i^s \in H^1\left(\Omega^s\right),
\left.v_i^s\right|_{\Gamma^s}=\left.v_i^f\right|_{\Gamma^s},
\left.v_i^f\right|_{\Gamma^D}=0  \right\}.
$$
Both the trial and test spaces for pressure in $\Omega^f$ are $L^2\left(\Omega^f\right)$, and both the trial and test spaces for pressure in $\Omega^s$ are $L^2\left(\Omega^s\right)$. We then perform the following symbolic operations:
$$
\left({\rm Eq.}(\ref{fluid1}),v_i^f\right)_{\Omega^f}-\left({\rm Eq.}(\ref{fluid2}),q^f\right)_{\Omega^f}
+\left({\rm Eq.}(\ref{solid1}),v_i^s\right)_{\Omega^s}-\left({\rm Eq.}(6^\prime),q^s\right)_{\Omega^s}.
$$
Integrating the stress terms by parts, using constitutive equation (\ref{fluid3}) and (\ref{solid3}) and boundary condition (\ref{NeummanBC}), the last operations give the following weak form of the FSI system. \\

Find $\left(u_i^f,u_i^s\right)\in W$, $p^f\in L^2\left(\Omega^f\right)$ and $p^s\in L^2\left(\Omega^s\right)$ such that
\begin{equation}\label{weakfluidsolid}
\begin{split}
& \rho^f\left( \frac{D{u_i^f}}{Dt}, v_i^f \right)_{\Omega^f}
+\left(\tau_{ij}^f, \frac{\partial v_i^f}{\partial x_j}\right)_{\Omega^f}
-\left(p^f,\frac{\partial v_j^f}{\partial x_j}\right)_{\Omega^f}
-\left(\frac{\partial u_j^f}{\partial x_j},q^f\right)_{\Omega^f}  \\
& \rho^s\left( \frac{D{u_i^s}}{Dt}, v_i^s \right)_{\Omega^s}
+\left(\tau_{ij}^s, \frac{\partial v_i^s}{\partial x_j}\right)_{\Omega^s}
-\left(p^s,\frac{\partial v_j^s}{\partial x_j}\right)_{\Omega^s}
-\left(\frac{\partial u_j^s}{\partial x_j},q^s\right)_{\Omega^s}  \\
&= \left(\bar{h}_i, v_i^f\right)_{\Gamma^N}
+\rho^f\left(g_i, v_i^f\right)_{\Omega^f}
+\rho^s\left(g_i, v_i^s\right)_{\Omega^s},
\end{split}
\end{equation}
$\forall \left(v_i^f,v_i^s\right) \in W_0$, $\forall q^f \in L^2\left({\Omega^f}\right)$ and  $\forall q^s \in L^2\left({\Omega^s}\right)$. \\
Note that the integrals on the interface (boundary forces) are also cancelled out using boundary condition (\ref{interfaceBC2}). This is not surprising because they are internal forces for the whole FSI system considered here.

We next extend the fluid velocity and pressure into solid domain by introducing 
$
u_i=\left \{ 
\begin{matrix}
{u_i^f \quad in \quad \Omega^f} \\
{u_i^s \quad in \quad \Omega^s} \\
\end{matrix}\right.
$
,
$
v_i=\left \{ 
\begin{matrix}
{v_i^f \quad in \quad \Omega^f} \\
{v_i^s \quad in \quad \Omega^s} \\
\end{matrix}\right.
$
,
$
p=\left \{ 
\begin{matrix}
{p^f \quad in \quad \Omega^f} \\
{p^s \quad in \quad \Omega^s} \\
\end{matrix}\right.
$
and
$
q=\left \{ 
\begin{matrix}
{q^f \quad in \quad \Omega^f} \\
{q^s \quad in \quad \Omega^s} \\
\end{matrix}\right.
$
, then extend the fluid computational domain from $\Omega^f$ to an augmented domain $\Omega$, and define a trial space for velocity in $\Omega$ as:
$$
\overline{W}=\left\{u_i:u_i \in H^1\left(\Omega\right), R\left(u_i\right)=u_i^s \in H^1\left(\Omega^s\right), \left.u_i\right|_{\Gamma^D}=\bar{u}_i \right\},
$$
with a corresponding test space for the velocity as:
$$
\overline{W}_0=\left\{v_i:v_i \in H^1\left(\Omega\right), R\left(v_i\right)=v_i^s \in H^1\left(\Omega^s\right), \left.v_i\right|_{\Gamma^D}=0 \right \},
$$
where $R\left(u_i\right)=\left.u_i\right|_{\Omega^s}$ is the restriction map. \\

Notice that $p^f$ and $p^s$ are not uniquely determined in (\ref{weakfluidsolid}). In fact, taking $p^f+c$ and $p^s+c$ instead of $p^f$ and $p^s$ respectively, the left-hand side of (\ref{weakfluidsolid}) does not change. This situation can be avoided by fixing the pressure at a selected point $\left(P_0\right)$ or by imposing the following constraint \cite{Boffi_2016}:

\begin{equation}
\int_{\Omega^f} p^fd\Omega+\int_{\Omega^s}p^sd\Omega=\int_\Omega pd\Omega=0.
\end{equation}
We shall use the former approach therefore define the trial space for pressure in $\Omega$ as:
$$
L_0^2\left(\Omega\right)=\left \{p:p \in L^2(\Omega), p\left|_{P_0}\right.=0 \right\}.
$$
The weak form of the FSI system in the augmented domain $\Omega$ can now be reformulated by rearranging equation (\ref{weakfluidsolid}) to yield the following formulation.

Find $u_i\in \overline{W}$ and $p\in L_0^2\left(\Omega\right)$ such that
\begin{equation}\label{weakaugmented}
\begin{split}
& \rho^f\left( \frac{D{u_i}}{Dt}, v_i \right)_\Omega
+\left(\tau_{ij}^f, \frac{\partial v_i}{\partial x_j}\right)_\Omega
-\left(p,\frac{\partial v_j}{\partial x_j}\right)_\Omega
-\left(\frac{\partial u_j}{\partial x_j}, q\right)_\Omega  \\
& +\left(\rho^s-\rho^f\right)\left( \frac{D{u_i}}{Dt}, v_i\right)_{\Omega^s}
+\left(\tau_{ij}^s-\tau_{ij}^f, \frac{\partial v_i}{\partial x_j}\right)_{\Omega^s} \\
&= \left(\bar{h}_i, v_i\right)_{\Gamma^N}
+\rho^f\left(g_i, v_i\right)_{\Omega}
+\left(\rho^s-\rho^f\right)\left(g_i, v_i\right)_{\Omega^s},
\end{split}
\end{equation}
$\forall v_i \in \overline{W}_0$ and $\forall q \in L^2\left({\Omega}\right)$.

{\bf Remark 2} The fluid deviatoric stress $\tau_{ij}^f$ is generally far smaller than the solid deviatoric stress $\tau_{ij}^s$, so we choose to neglect the fluid deviatoric stress $\tau_{ij}^f$ in $\Omega^s$ in what follows. Note that the classical IFEM neglects the whole fluid stress $\sigma_{ij}^f$ when computing the FSI force \cite{zhang2004immersed}. An equivalent way of interpreting neglecting $\tau_{ij}^f$ in $\Omega^s$ is to view the solid as being slightly visco-elastic, having the same viscosity as the fluid.

{\bf Remark 3} We treat the solid as a freely moving object in a fluid, so $u_i^s,v_i^s\in H^1\left(\Omega^s\right)$ without any boundary constraints in the definition of $\overline{W}$ and $\overline{W}_0$ respectively. Physically, however, if part of solid boundary is fixed, this fixed boundary can also be regarded as a fixed fluid boundary and implemented as a zero velocity condition in the fluid domain, hence the solid still can be treated as if it were freely moving. Furthermore, the interface boundary condition (\ref{interfaceBC1}) is automatically built into the solution because we use an augmented solution space $\overline{W}$ which requires $\left.u_i\right|_{\Omega^s}=u_i^s$.

\section{Computational scheme}
The integrals in equation (\ref{weakaugmented}) are carried out in two different domains as illustrated in Figure \ref{fig1}. We use an Eulerian mesh to represent $\Omega$ and an updated Lagrangian mesh to represent $\Omega^s$, therefore the total time derivatives in these two different domains have different expressions, i.e: 
\begin{equation}\label{derivativef}
\frac{Du_i}{Dt}=\frac{\partial u_i}{\partial t}+u_j\frac{\partial u_i}{\partial x_j} \quad in \quad \Omega,
\end{equation}
and
\begin{equation}\label{derivatives}
\frac{Du_i^s}{Dt}=\frac{\partial u_i^s}{\partial t} \quad in \quad \Omega^s.
\end{equation}

Standard FEM isoparametric interpolation may be used to transfer data between the two meshes. Firstly, based on the above two equations (\ref{derivativef}) and (\ref{derivatives}), we discretize (\ref{weakaugmented}) in time using a backward finite difference. Then omiting the superscript $n+1$, showing the solution is at the end of the time step, for convenience, we obtain:
\begin{equation}\label{weak2}
\begin{split}
& \rho^f\left( \frac{u_i-u_i^n}{\Delta t}+u_j\frac{\partial u_i}{\partial x_j}, v_i \right)_\Omega
+\left(\tau_{ij}^f, \frac{\partial v_i}{\partial x_j}\right)_\Omega
-\left(p,\frac{\partial v_j}{\partial x_j}\right)_\Omega-\left(\frac{\partial u_j}{\partial x_j}, q\right)_\Omega  \\
& +\left(\rho^s-\rho^f\right)\left( \frac{u_i-u_i^n}{\Delta t}, v_i\right)_{\Omega^s}
+\left(\tau_{ij}^s, \frac{\partial v_i}{\partial x_j}\right)_{\Omega^s} \\
&= \left(\bar{h}_i, v_i\right)_{\Gamma^N}
+\rho^f\left(g_i, v_i\right)_{\Omega}
+\left(\rho^s-\rho^f\right)\left(g_i, v_i\right)_{\Omega^s}.
\end{split}
\end{equation}

Using the splitting method of \cite[Chapter 3]{Zienkiewic2014}, equation (\ref{weak2}) can be expressed in the following two steps.

(1) Convection step:
\begin{equation}\label{weakconvection}
\rho^f\left(\frac{u_i^*-u_i^n}{\Delta t}+u_j^*\frac{\partial u_i^*}{\partial x_j}, v_i\right)_\Omega=0;
\end{equation}

(2) Diffusion step:
\begin{equation}\label{weakdiffusion}
\begin{split}
& \rho^f\left( \frac{u_i-u_i^*}{\Delta t}, v_i \right)_\Omega
+\left(\tau_{ij}^f, \frac{\partial v_i}{\partial x_j}\right)_\Omega
-\left(p,\frac{\partial v_j}{\partial x_j}\right)_\Omega-\left(\frac{\partial u_j}{\partial x_j}, q\right)_\Omega  \\
& +\left(\rho^s-\rho^f\right)\left( \frac{u_i-u_i^n}{\Delta t}, v_i \right)_{\Omega^s}
+\left(\tau_{ij}^s, \frac{\partial v_i}{\partial x_j}\right)_{\Omega^s} \\
&= \left(\bar{h}_i, v_i\right)_{\Gamma^N}
+\rho^f\left(g_i, v_i\right)_{\Omega}
+\left(\rho^s-\rho^f\right)\left(g_i, v_i\right)_{\Omega^s}.
\end{split}
\end{equation}
The treatment of the above two steps is described separately in the following subsections.

\subsection{Linearization of the convection step}
In this section, two methods are introduced to treat the convection equation: the implicit Least-squares method and the explict Taylor-Galerkin method, both of which can be used in the framework of our UFEM scheme. Some numerical results for comparison between these two methods are discussed subsequently in section 5.

\subsubsection{Implicit Least-squares method}
It is possible to linearize (\ref{weakconvection}) using the value of $u_i$ from the last time step:
\begin{equation}\label{linearizationofconvection}
u_j^*\frac{\partial u_i^*}{\partial x_j}\approx  u_j^*\frac{\partial u_i^n}{\partial x_j}
+u_j^n\frac{\partial u_i^*}{\partial x_j}-u_j^n\frac{\partial u_i^n}{\partial x_j}.
\end{equation}
Substituting (\ref{linearizationofconvection}) into equation (\ref{weakconvection}) gives,
\begin{equation}
\left(u_i^*+\Delta t\left(u_j^*\frac{\partial u_i^n}{\partial x_j}+u_j^n\frac{\partial u_i^*}{\partial x_j}\right), v_i\right)_\Omega
=\left(u_i^n+\Delta t u_j^n\frac{\partial u_i^n}{\partial x_j}, v_i\right)_\Omega.
\end{equation}
For the Least-squares method \cite{bochev2009least}, we may choose the test function in the following form:
\begin{equation}
v_i=L\left(w_i\right)=w_i+\Delta t\left(w_j\frac{\partial u_i^n}{\partial x_j}+u_j^n\frac{\partial w_i}{\partial x_j}\right),
\end{equation}
where $w_i \in \overline{W}_0$. In such a case, the weak form of (\ref{weakconvection}) is:
\begin{equation}\label{lastweakofconvection}
\left(L\left(u_i^*\right), L\left(w_i\right)\right)_\Omega
=\left(u_i^n+\Delta t u_j^n\frac{\partial u_i^n}{\partial x_j}, L\left(w_i\right)\right)_\Omega.
\end{equation}
In our UFEM a standard biquadratic finite element space is used to discretize equation (\ref{lastweakofconvection}) directly, although other spaces could be used.

\subsubsection{Explicit Taylor-Galerkin method}
It is also possible to linearize equation (\ref{weakconvection}) as:
\begin{equation}\label{cbsconvection1}
\left(\frac{u_i^*-u_i^n}{\Delta t}+\frac{1}{2}u_j^n\frac{\partial}{\partial x_j}\left(u_i^*+u_i^n\right), v_i\right)_\Omega=0,
\end{equation}
or
\begin{equation}\label{cbsconvection2}
\left(\frac{u_i^*-u_i^n}{\Delta t}+u_j^n\frac{\partial u_i^n}{\partial x_j}, v_i\right)_\Omega=0.
\end{equation}
Re-write (\ref{cbsconvection2}) as:
\begin{equation}\label{cbsconvection-strong-form}
u_i^*=u_i^n-{\Delta t}u_j^n\frac{\partial u_i^n}{\partial x_j},
\end{equation}
and substitute (\ref{cbsconvection-strong-form}) into equation (\ref{cbsconvection1}), we have
\begin{equation}\label{cbs-strong-into}
\left(\frac{u_i^*-u_i^n}{\Delta t}+u_j^n\frac{\partial u_i^n}{\partial x_j}
-\frac{\Delta t}{2}u_j^n\frac{\partial}{\partial x_j}\left(u_k^n\frac{\partial u_i^n}{\partial x_k}\right), v_i\right)_\Omega=0.
\end{equation}
Notice that a second order derivative exists in the last equation. In practice, one does not need to calculate the second order derivative, instead, Integration by parts may be used to reduce the order:
\begin{equation}\label{cbs-integrate-by-parts}
\left(\frac{\partial}{\partial x_j}\left(u_k\frac{\partial u_i}{\partial x_k}\right), v_i\right)_\Omega=
\left(u_k\frac{\partial u_i}{\partial x_k}, v_i\right)_{\Gamma^N}
-\left(u_k\frac{\partial u_i}{\partial x_k}, \frac{\partial v_i}{\partial x_j}\right)_\Omega.
\end{equation}
The boundary integral in the last equation can be neglected if $u_i$ is the solution of the previous diffusion step, which means no convection exists on the boundary after the diffusion step. Using (\ref{cbs-integrate-by-parts}), equation (\ref{cbs-strong-into}) may be approximated as:
\begin{equation}\label{cbs-last2}
\left(\frac{u_i^*-u_i^n}{\Delta t}+u_j^n\frac{\partial u_i^n}{\partial x_j}, v_i\right)_\Omega
=-\frac{\Delta t}{2}\left(u_k^n\frac{\partial u_i^n}{\partial x_k}, u_j^n\frac{\partial v_i}{\partial x_j}\right)_\Omega.
\end{equation}

At last the weak form of the Taylor-Galerkin method \cite[Chapter 2]{Zienkiewic2014} can be expressed, by rearranging the last equation, as:
\begin{equation}\label{cbs-last}
\left(u_i^*,v_i\right)_\Omega=
\left(u_i^n-\Delta t u_j^n\frac{\partial u_i^n}{\partial x_j},v_i\right)_\Omega
-\frac{\Delta t^2}{2}\left(u_k^n\frac{\partial u_i^n}{\partial x_k}, u_j^n\frac{\partial v_i}{\partial x_j}\right)_\Omega.
\end{equation}
This Taylor-Galerkin method is explicit, however a small time step is usually needed to keep the scheme stable.

\subsection{Linearization of the diffusion step}
In both the above and the following context, the derivative $\frac{\partial}{\partial x_i}$ on the updated solid mesh is computed at the current known coordinates $x_i^n$, that is to say $\frac{\partial}{\partial x_i}=\frac{\partial}{\partial x_i^n}$. Furthermore, $\tau_{ij}^s$ in equations (\ref{weakdiffusion}), has a nonlinear relationship with $x_i$, i.e.:
\begin{equation}
\tau_{ij}^s=\left(\tau_{ij}^s\right)^{n+1}=\mu^s\left(\frac{\partial x_i^{n+1}}{\partial X_k}
\frac{\partial x_j^{n+1}}{\partial X_k}-\delta_{ij}\right).
\end{equation}
Using a chain rule, the last equation can also be expressed as:
\begin{equation}
\begin{split}
&\left(\tau_{ij}^s\right)^{n+1}=\mu^s\frac{\partial x_i^{n+1}}{\partial x_k^n}\frac{\partial x_k^n}{\partial X_m}
\frac{\partial x_l^n}{\partial X_m}\frac{\partial x_j^{n+1}}{\partial x_l^n} -\mu^s\delta_{ij}\\
&+\mu^s\frac{\partial x_i^{n+1}}{\partial x_k^n}\frac{\partial x_j^{n+1}}{\partial x_k^n} 
-\mu^s\frac{\partial x_i^{n+1}}{\partial x_k^n}{\delta_{kl}}\frac{\partial x_j^{n+1}}{\partial x_l^n}
\end{split}
\end{equation}
or
\begin{equation}
\begin{split}
&\left(\tau_{ij}^s\right)^{n+1}=\mu^s\left(\frac{\partial x_i^{n+1}}{\partial x_k^n}
\frac{\partial x_j^{n+1}}{\partial x_k^n}-\delta_{ij}\right) \\
&+\mu^s\frac{\partial x_i^{n+1}}{\partial x_k^n}\left(\frac{\partial x_k^n}{\partial X_m}
\frac{\partial x_l^n}{\partial X_m}-{\delta_{kl}}
\right)\frac{\partial x_j^{n+1}}{\partial x_l^n}
\end{split}
,
\end{equation}
and then $\left(\tau_{ij}^s\right)^{n+1}$ can be expressed by the current coordinate $x_i^n$ as follows:
\begin{equation}
\begin{split}
&\left(\tau_{ij}^s\right)^{n+1}=\mu^s\left(\frac{\partial x_i^{n+1}}{\partial x_k^n}
\frac{\partial x_j^{n+1}}{\partial x_k^n}-\delta_{ij}\right)
+\frac{\partial x_i^{n+1}}{\partial x_k^n}\left(\tau_{kl}^s\right)^n
\frac{\partial x_j^{n+1}}{\partial x_l^n}
\end{split}
.
\end{equation}
Using $x_i^{n+1}-x_i^n=u_i^{n+1}\Delta t$ which is the displacement at the current step, the last equation can also be expressed as:
\begin{equation}\label{solidstress}
\begin{split}
&\left(\tau_{ij}^s\right)^{n+1}=\mu^s \Delta t\left(\frac{\partial u_i^{n+1}}{\partial x_j^n}+\frac{\partial u_j^{n+1}}{\partial x_i^n}+\Delta t\frac{\partial u_i^{n+1}}{\partial x_k^n}\frac{\partial u_j^{n+1}}{\partial x_k^n}\right)+\left(\tau_{ij}^s\right)^n   \\
&+\Delta t^2\frac{\partial u_i^{n+1}}{\partial x_k^n}\left(\tau_{kl}^s\right)^n\frac{\partial u_j^{n+1}}{\partial x_l^n}
+\Delta t\frac{\partial u_i^{n+1}}{\partial x_k^n}\left(\tau_{kj}^s\right)^n
+\Delta t\left(\tau_{il}^s\right)^n\frac{\partial u_j^{n+1}}{\partial x_l^n}.
\end{split}
\end{equation}
There are two nonlinear terms in the last equation. Using a Newton method, they can be linearized as follows.
\begin{equation}\label{linearization1}
\frac{\partial u_i^{n+1}}{\partial x_k^n}\frac{\partial u_j^{n+1}}{\partial x_k^n}
=\frac{\partial u_i^{n+1}}{\partial x_k^n}\frac{\partial u_j^{n}}{\partial x_k^n}
+\frac{\partial u_i^{n}}{\partial x_k^n}\frac{\partial u_j^{n+1}}{\partial x_k^n}
-\frac{\partial u_i^{n}}{\partial x_k^n}\frac{\partial u_j^{n}}{\partial x_k^n}
\end{equation}
and
\begin{equation}\label{linearization2}
\begin{split}
&\frac{\partial u_i^{n+1}}{\partial x_k^n}\left(\tau_{kl}^s\right)^n\frac{\partial u_j^{n+1}}{\partial x_l^n}
=\frac{\partial u_i^{n+1}}{\partial x_k^n}\left(\tau_{kl}^s\right)^n\frac{\partial u_j^{n}}{\partial x_l^n} \\
&+\frac{\partial u_i^{n}}{\partial x_k^n}\left(\tau_{kl}^s\right)^n\frac{\partial u_j^{n+1}}{\partial x_l^n}
-\frac{\partial u_i^{n}}{\partial x_k^n}\left(\tau_{kl}^s\right)^n\frac{\partial u_j^{n}}{\partial x_l^n}.
\end{split}
\end{equation}

Substituting (\ref{solidstress})-(\ref{linearization2}) into (\ref{weakdiffusion}) and dropping off the superscripts $n+1$ of $u_i^{n+1}$ for notation convenience, this may be expressed as:
\begin{equation}\label{lastbigequation}
\begin{split}
&\rho^f\left(\frac{u_i-u_i^*}{\Delta t}, v_i\right)_\Omega+\left(\rho^s-\rho^f\right)\left(\frac{u_i^s-\left(u_i^s\right)^n}{\Delta t}, v_i\right)_{\Omega^s} \\
&+\mu^f\left(\frac{\partial u_i}{\partial x_j}+\frac{\partial u_j}{\partial x_i}, \frac{\partial v_i}{\partial x_j}\right)_\Omega
-\left(p, \frac{\partial v_j}{\partial x_j}\right)_\Omega-\left(\frac{\partial u_j}{\partial x_j}, q\right)_\Omega \\
&+\mu^s\Delta t\left(\frac{\partial u_i}{\partial x_j}+\frac{\partial u_j}{\partial x_i}
+\Delta t\frac{\partial u_i}{\partial x_k}\frac{\partial u_j^n}{\partial x_k}
+\Delta t\frac{\partial u_i^n}{\partial x_k}\frac{\partial u_j}{\partial x_k}, 
\frac{\partial v_i}{\partial x_j}  \right)_{\Omega^s} \\
&+\Delta t^2\left(\frac{\partial u_i}{\partial x_k}\left(\tau_{kl}^s\right)^n\frac{\partial u_j^n}{\partial x_l}  
+\frac{\partial u_i^n}{\partial x_k}\left(\tau_{kl}^s\right)^n\frac{\partial u_j}{\partial x_l}, 
\frac{\partial v_i}{\partial x_j}\right)_{\Omega^s} \\
&+\Delta t\left(\frac{\partial u_i}{\partial x_k}\left(\tau_{kj}^s\right)^n+\left(\tau_{il}^s\right)^n\frac{\partial u_j}{\partial x_l}, \frac{\partial v_i}{\partial x_j}\right)_{\Omega^s} \\
&=\left(\bar{h}_i, v_i\right)_{\Gamma^N}+\rho^f\left(g_i, v_i\right)_\Omega+\left(\rho^s-\rho^f\right)\left(g_i, v_i\right)_{\Omega^s} \\
&+\left(\mu^s\Delta t^2\frac{\partial u_i^n}{\partial x_k}\frac{\partial u_j^n}{\partial x_k}
+\Delta t^2\frac{\partial u_i^n}{\partial x_k}\left(\tau_{kl}^s\right)^n\frac{\partial u_j^n}{\partial x_l}-\left(\tau_{ij}^s\right)^n, \frac{\partial v_i}{\partial x_j}
\right)_{\Omega^s}.
\end{split}
\end{equation}

The spatial discretization of the above linearized weak form will be discussed in the following section.

\subsection{Discretization in space} 
In the 2D case, which is considered in the remainder of this paper, a standard Taylor-Hood element Q2Q1 (9-node biquadratic quadrilateral for velocity and 4-node bilinear quadrilateral for pressure) is used to discretize in space. We first discretize the domain $\Omega$ to get $\Omega^h$, then define finite dimensional subspaces of $\overline{W}$ and $\overline{W}_0$ as follows.

The solution space for each component of velocity:
$$
\overline{W}^h=\left\{u_i^h:u_i^h \in H^{1h}\left(\Omega^h\right), R^h\left(u_i^h\right)=u_i^{sh}, \left.u_i^h\right|_{\Gamma^D}=\bar{u}_i^h \right\},
$$
whilst test space for velocity is
$$
\overline{W}_0^h=\left\{v_i^h:v_i^h \in H^{1h}\left(\Omega^h\right), R^h\left(v_i^h\right)=v_i^{sh}, \left.v_i^h\right|_{\Gamma^D}=0\right\}.
$$

We also discretize the domain $\Omega^s$ to get $\Omega^{sh}$, and both the discretized trial space and test space on the solid domain are $H^{1h}\left(\Omega^{sh}\right)$ based on the discussion of Remark 3.

The solution and test spaces for pressure are $L_0^{2h}\left(\Omega^h\right)$ and $L^{2h}\left(\Omega^h\right)$ respectively, which represent the finite dimensional subspaces of $L_0^2\left(\Omega\right)$ and $L^2\left(\Omega\right)$, respectively, based on continuous piecewise bilinear functions. $H^{1h}\left(\Omega^h\right)$ $\left(H^{1h}\left(\Omega^{sh}\right)\right)$ represents the finite dimensional subspace of $H^1\left(\Omega\right)$ $\left(H^1\left(\Omega^s\right)\right)$ based upon continuous piecewise biquadratic functions. Then equation (\ref{lastbigequation}) can be discretized as:
\begin{equation}\label{last-h-equation}
\begin{split}
&\rho^f\left(\frac{u_i^h-u_i^{*h}}{\Delta t}, v_i^h\right)_{\Omega^h}+\left(\rho^s-\rho^f\right)\left(\frac{u_i^{sh}-\left(u_i^{sh}\right)^n}{\Delta t}, v_i^{sh}\right)_{\Omega^{sh}} \\
&+\mu^f\left(\frac{\partial u_i^h}{\partial x_j}+\frac{\partial u_j^h}{\partial x_i}, \frac{\partial v_i^h}{\partial x_j}\right)_{\Omega^h}
-\left(p^h, \frac{\partial v_j^h}{\partial x_j}\right)_{\Omega^h}-\left(\frac{\partial u_j^h}{\partial x_j}, q^h\right)_{\Omega^h} \\
&+\mu^s\Delta t\left(\frac{\partial u_i^{sh}}{\partial x_j}+\frac{\partial u_j^{sh}}{\partial x_i}
+\Delta t\frac{\partial u_i^{sh}}{\partial x_k}\frac{\partial u_j^n}{\partial x_k}
+\Delta t\frac{\partial u_i^n}{\partial x_k}\frac{\partial u_j^{sh}}{\partial x_k}, 
\frac{\partial v_i^{sh}}{\partial x_j}  \right)_{\Omega^{sh}} \\
&+\Delta t^2\left(\frac{\partial u_i^{sh}}{\partial x_k}\left(\tau_{kl}^s\right)^n\frac{\partial u_j^n}{\partial x_l}  
+\frac{\partial u_i^n}{\partial x_k}\left(\tau_{kl}^s\right)^n\frac{\partial u_j^{sh}}{\partial x_l}, 
\frac{\partial v_i^{sh}}{\partial x_j}\right)_{\Omega^{sh}} \\
&+\Delta t\left(\frac{\partial u_i^{sh}}{\partial x_k}\left(\tau_{kj}^s\right)^n+\left(\tau_{il}^s\right)^n\frac{\partial u_j^{sh}}{\partial x_l}, \frac{\partial v_i^{sh}}{\partial x_j}\right)_{\Omega^{sh}} \\
&=\left(\bar{h}_i, v_i^h\right)_{\Gamma^{Nh}}+\rho^f\left(g_i, v_i^h\right)_{\Omega^h}+\left(\rho^s-\rho^f\right)\left(g_i, v_i^{sh}\right)_{\Omega^{sh}} \\
&+\left(\mu^s\Delta t^2\frac{\partial u_i^n}{\partial x_k}\frac{\partial u_j^n}{\partial x_k}
+\Delta t^2\frac{\partial u_i^n}{\partial x_k}\left(\tau_{kl}^s\right)^n\frac{\partial u_j^n}{\partial x_l}-\left(\tau_{ij}^s\right)^n, \frac{\partial v_i^{sh}}{\partial x_j}
\right)_{\Omega^{sh}}.
\end{split}
\end{equation}
Notice that in the continuous space $\overline{W}$, we have the restriction map $R\left(u_i\right)=\left.u_i\right|_{\Omega^s}=u_i^s$, while in the discretized space $\overline{W}^h$, we use the standard FEM isoparametric transformation $R^h$ to represent the map, i.e.
\begin{equation}\label{interpolaltion}
u_i^{sh}=R^h\left(u_i^h\right),
\end{equation}
where subscript $i$ denotes the velocity components in each space dimension.

Let $\widetilde{\bf u}_i=\left(\widetilde{u}_{i1}, \widetilde{u}_{i2}\cdots\widetilde{u}_{iN^f} \right)^{\rm T}$ and $\widetilde{\bf u}_i^s=\left(\widetilde{u}_{i1}^s, \widetilde{u}_{i2}^s\cdots\widetilde{u}_{iN^s}^s \right)^{\rm T}$ denote the $i^{th}$ components of the nodal velocity vectors on the fluid and solid meshes respectively, and $\boldsymbol{\varphi}=\left(\varphi_1, \varphi_2 \cdots \varphi_{N^f}\right)^{\rm T}$ and $\boldsymbol{\varphi}^s=\left(\varphi_1^s, \varphi_2^s \cdots \varphi_{N^s}^s\right)^{\rm T}$ denote the vector of velocity basis functions on the fluid and solid meshes respectively, where $N^f$ and $N^s$ are the number of nodes of fluid and solid mesh respectively. Then equation (\ref{interpolaltion}) can be expressed as:
\begin{equation}
\widetilde{u}_{ik}^s\varphi_k^s=R^h\left(\widetilde{u}_{ik}\varphi_k\right).
\end{equation}
The FEM isoparametric transformation defines $R^h$ from $\widetilde{\bf u}_i$ to $\widetilde{\bf u}_i^s$ as follows:
\begin{equation}
\widetilde{u}_{ik}^s=R^h\left(\widetilde{u}_{il}\right)=\widetilde{u}_{il}R_{lk},
\end{equation}
where $R_{lk}=\varphi_l\left({\bf x}_k\right)$, ${\bf x}_k\left(k=1, 2\cdots N^s\right)$ is the current coordinate of the $k^{th}$ node on the solid mesh. Therefore,
\begin{equation}\label{shapfirst}
u_i^{sh}=\widetilde{u}_{ik}^s\varphi_k^s=\widetilde{u}_{il}R_{lk}\varphi_k^s.
\end{equation}

For velocity test functions, we similarly have
\begin{equation}
v_i^{sh}=\widetilde{v}_{ik}^s\varphi_k^s=\widetilde{v}_{il}R_{lk}\varphi_k^s,
\end{equation}
where $\widetilde{\bf v}_i=\left(\widetilde{v}_{i1}, \widetilde{v}_{i2}\cdots\widetilde{v}_{iN^f} \right)^{\rm T}$ is an arbitrary nodal velocity (virtual velocity) vector on the fluid mesh, which satisfies the homogeneous Dirichlet boundary condition.

On the fluid mesh, velocity and pressure can also be expressed as follows:
\begin{equation}
u_i^h=\widetilde{u}_{ik}\varphi_k,
\end{equation}
\begin{equation}
v_i^h=\widetilde{v}_{ik}\varphi_k,
\end{equation}
\begin{equation}
p^h=\widetilde{p}_k\psi_k,
\end{equation}
\begin{equation}\label{shaplast}
q^h=\widetilde{q}_k\psi_k,
\end{equation}
where $\boldsymbol{\psi}=\left(\psi_1, \psi_2 \cdots\ \psi_{N^p}\right)^{\rm T}$ is the vector of pressure basis functions, $\widetilde{\bf p}=\left(\widetilde{p}_1, \widetilde{p}_2,\cdots\ \widetilde{p}_{N^p}\right)^{\rm T}$ is the nodal pressure vector, and $\widetilde{\bf q}=\left(\widetilde{q}_1, \widetilde{q}_2,\cdots\ \widetilde{q}_{N^p}\right)^{\rm T}$ is an arbitrary nodal pressure vector. $N^p$ denotes the number of nodes on the fluid mesh at which only pressure is defined.

Substituting (\ref{shapfirst})-(\ref{shaplast}) into (\ref{last-h-equation}), we have 
\begin{equation}\label{lastbigequationofdiscretization}
\begin{split}
&\rho^f\left(\frac{\widetilde{u}_{ik}-\widetilde{u}_{ik}^*}{\Delta t}\varphi_k, \widetilde{v}_{im}\varphi_m\right)_{\Omega^h} \\
&+\left(\rho^s-\rho^f\right)\left(\frac{\widetilde{u}_{il}-\widetilde{u}_{il}^n}{\Delta t}R_{lk}\varphi_k^s, \widetilde{v}_{ir}R_{rm}\varphi_m^s\right)_{\Omega^{sh}} \\
&+\mu^f\left(\widetilde{u}_{ik}\frac{\partial \varphi_k}{\partial x_j}+\widetilde{u}_{jk}\frac{\partial \varphi_k}{\partial x_i}, \widetilde{v}_{im}\frac{\partial \varphi_m}{\partial x_j}\right)_{\Omega^h} \\
&-\left(\widetilde{p}_k\psi_k, \widetilde{v}_{jm}\frac{\partial \varphi_m}{\partial x_j}\right)_{\Omega^h}-\left(\widetilde{u}_{jk}\frac{\partial \varphi_k}{\partial x_j}, \widetilde{q}_m\psi_m\right)_{\Omega^h} \\
&+\mu^s\Delta t\left(\widetilde{u}_{il}R_{lk}\frac{\partial \varphi_k^s}{\partial x_j}
+\widetilde{u}_{jl}R_{lk}\frac{\partial \varphi_k^s}{\partial x_i}, 
\widetilde{v}_{ir}R_{rm}\frac{\partial \varphi_m^s}{\partial x_j}  \right)_{\Omega^{sh}} \\
&+\mu^s\Delta t^2\left(\widetilde{u}_{ia}R_{ab}\frac{\partial \varphi_b^s}{\partial x_k}\frac{\partial u_j^n}{\partial x_k}
+\widetilde{u}_{ja}R_{ab}\frac{\partial u_i^n}{\partial x_k}\frac{\partial \varphi_b^s}{\partial x_k}, 
\widetilde{v}_{ir}R_{rm}\frac{\partial \varphi_m^s}{\partial x_j}  \right)_{\Omega^{sh}} \\
&+\Delta t^2\left(\widetilde{u}_{ia}R_{ab}\frac{\partial \varphi_b^s}{\partial x_k}\left(\tau_{kl}^s\right)^n\frac{\partial u_j^n}{\partial x_l}, 
\widetilde{v}_{ir}R_{rm}\frac{\partial \varphi_m^s}{\partial x_j}\right)_{\Omega^{sh}} \\
&+\Delta t^2\left(\widetilde{u}_{ja}R_{ab}\frac{\partial u_i^n}{\partial x_k}\left(\tau_{kl}^s\right)^n\frac{\partial \varphi_b^s}{\partial x_l}, 
\widetilde{v}_{ir}R_{rm}\frac{\partial \varphi_m^s}{\partial x_j}\right)_{\Omega^{sh}} \\
&+\Delta t\left(\widetilde{u}_{ia}R_{ab}\frac{\partial \varphi_b^s}{\partial x_k}\left(\tau_{kj}^s\right)^n
+\left(\tau_{il}^s\right)^n \widetilde{u}_{ja}R_{ab}\frac{\partial \varphi_b^s}{\partial x_l}, \widetilde{v}_{ir}R_{rm}\frac{\partial \varphi_m^s}{\partial x_j}\right)_{\Omega^{sh}} \\
&=\left(\bar{h}_i, \widetilde{v}_{im}\varphi_m\right)_{\Gamma^{Nh}}+\rho^f\left(g_i, \widetilde{v}_{im}\varphi_m\right)_{\Omega^h}+\left(\rho^s-\rho^f\right)\left(g_i, \widetilde{v}_{ir}R_{rm}\varphi_m^s\right)_{\Omega^{sh}} \\
&+\left(\mu^s\Delta t^2\frac{\partial u_i^n}{\partial x_k}\frac{\partial u_j^n}{\partial x_k}
+\Delta t^2\frac{\partial u_i^n}{\partial x_k}\left(\tau_{kl}^s\right)^n\frac{\partial u_j^n}{\partial x_l}-\left(\tau_{ij}^s\right)^n, \widetilde{v}_{ir}R_{rm}\frac{\partial \varphi_m^s}{\partial x_j}
\right)_{\Omega^{sh}}.
\end{split}
\end{equation}
Let $\widetilde{\bf u}=\left(\widetilde{\bf u}_1^{\rm T}, \widetilde{\bf u}_2^{\rm T}\right)^{\rm T}$ and $\widetilde{\bf v}=\left(\widetilde{\bf v}_1^{\rm T}, \widetilde{\bf v}_2^{\rm T}\right)^{\rm T}$, we then express (\ref{lastbigequationofdiscretization}) in the following matrix form:
\begin{equation}\label{initialenergyform}
\begin{split}
&\widetilde{\bf v}^{\rm T}{\bf M}\frac{\widetilde{\bf u}-\widetilde{\bf u}^*}{\Delta t}
+\widetilde{\bf v}^{\rm T}{\bf D}^{\rm T}{\bf M}^s{\bf D}\frac{\widetilde{\bf u}-\widetilde{\bf u}^n}{\Delta t} \\
&+\widetilde{\bf v}^{\rm T}{\bf K}\widetilde{\bf u}
+\widetilde{\bf v}^{\rm T}{\bf B}\widetilde{\bf p}+\widetilde{\bf q}^{\rm T}{\bf B}^{\rm T}\widetilde{\bf u}
+\widetilde{\bf v}^{\rm T}{\bf D}^{\rm T}{\bf K}^s{\bf D}\widetilde{\bf u} \\
&=\widetilde{\bf v}^{\rm T}{\bf f}+\widetilde{\bf v}^{\rm T}{\bf D}^{\rm T}{\bf f}^s,
\end{split}
\end{equation}
or
\begin{equation}\label{energyformoflinearalgbraic}
\begin{pmatrix}
\widetilde{\bf v}^{\rm T}, \widetilde{\bf q}^{\rm T}
\end{pmatrix}
\begin{bmatrix}
{\bf A} & {\bf B} \\
{\bf B}^{\rm T} &{\bf 0}
\end{bmatrix}
\begin{pmatrix}
\widetilde{\bf u} \\
\widetilde{\bf p}
\end{pmatrix}
=
\begin{pmatrix}
\widetilde{\bf v}^{\rm T}, \widetilde{\bf q}^{\rm T}
\end{pmatrix}
\begin{pmatrix}
\widetilde{\bf b} \\
{\bf 0}
\end{pmatrix}
,
\end{equation}
where
$$
{\bf A}={\bf M}/\Delta t+{\bf K}+{\bf D}^{\rm T}\left({\bf M}^s/\Delta t+{\bf K}^s\right){\bf D}
$$
and
$$
{\bf b}={\bf f}+{\bf D}^{\rm T}{\bf f}^s+{\bf M}\widetilde{\bf u}^*/\Delta t+{\bf D}^{\rm T}{\bf M}^s{\bf D}\widetilde{\bf u}^n/\Delta t.
$$
The matrix
\begin{equation}
{\bf M}=\rho^f
\begin{bmatrix}
{\bf M}_{11} & {} \\
{} & {\bf M}_{22} \\
\end{bmatrix}
\end{equation}
is the velocity mass matrix of the fluid, where
$$
\left({\bf M}_{11}\right)_{km}=\left({\bf M}_{22}\right)_{km}=\left(\varphi_k,\varphi_m\right)_{\Omega^h},\left(k,m=1,2,\cdots N^f\right).
$$
The matrix
\begin{equation}
{\bf M}^s=\left(\rho^s-\rho^f\right)
\begin{bmatrix}
{\bf M}_{11}^s & {} \\
{} & {\bf M}_{22}^s \\
\end{bmatrix}
\end{equation}
is the velocity mass matrix of the solid, where
$$
\left({\bf M}_{11}^s\right)_{km}=\left({\bf M}_{22}^s\right)_{km}=\left(\varphi_k^s,\varphi_m^s\right)_{\Omega^{sh}}, \left(k,m=1,2,\cdots N^s\right).
$$
$\bf K$ is the stiffness matrix of the fluid:
\begin{equation}
{\bf K}=\mu^f
\begin{bmatrix}
{{\bf K}_{11}} & {{\bf K}_{12}} \\
{{\bf K}_{21}} & {{\bf K}_{22}} \\
\end{bmatrix}
,
\end{equation}
where
$$
\left({\bf K}_{11}\right)_{km}=2\left(\frac{\partial \varphi_k}{\partial x_1}, \frac{\partial \varphi_m}{\partial x_1}\right)_{\Omega^h}+\left(\frac{\partial \varphi_k}{\partial x_2}, \frac{\partial \varphi_m}{\partial x_2}\right)_{\Omega^h},
$$
$$
\left({\bf K}_{22}\right)_{km}=2\left(\frac{\partial \varphi_k}{\partial x_2}, \frac{\partial \varphi_m}{\partial x_2}\right)_{\Omega^h}+\left(\frac{\partial \varphi_k}{\partial x_1}, \frac{\partial \varphi_m}{\partial x_1}\right)_{\Omega^h},
$$
$$
\left({\bf K}_{12}\right)_{km}=\left(\frac{\partial \varphi_k}{\partial x_1}, \frac{\partial \varphi_m}{\partial x_2}\right)_{\Omega^h},
$$
$$
\left({\bf K}_{21}\right)_{km}=\left({\bf K}_{12}\right)_{mk}=\left(\frac{\partial \varphi_k}{\partial x_2}, \frac{\partial \varphi_m}{\partial x_1}\right)_{\Omega^h},
$$
and $k,m=1,2,\cdots N^f$. \\
${\bf K}^s$  is the stiffness matrix of the solid:
\begin{equation}
{\bf K}^s=
\begin{bmatrix}
{{\bf K}_{11}^s} & {{\bf K}_{12}^s} \\
{\bf K}_{21}^s & {{\bf K}_{22}^s} \\
\end{bmatrix},
\end{equation}
where 
\begin{equation*}
\begin{split}
&\left({\bf K}_{11}^s\right)_{bm}=\mu^s \Delta t2\left(\frac{\partial \varphi_b^s}{\partial x_1}, \frac{\partial \varphi_m^s}{\partial x_1}\right)_{\Omega^{sh}}+\mu^s \Delta t\left(\frac{\partial \varphi_b^s}{\partial x_2},
\frac{\partial \varphi_m^s}{\partial x_2}\right)_{\Omega^{sh}} \\
& +2\mu^s \Delta t^2\left(\frac{\partial \varphi_b^s}{\partial x_k}\frac{\partial u_1^n}{\partial x_k}, \frac{\partial \varphi_m^s}{\partial x_1}\right)_{\Omega^{sh}}+\mu^s \Delta t^2\left(\frac{\partial \varphi_b^s}{\partial x_k}\frac{\partial u_2^n}{\partial x_k},
\frac{\partial \varphi_m^s}{\partial x_2}\right)_{\Omega^{sh}} \\
&+ 2\Delta t^2\left(\frac{\partial \varphi_b^s}{\partial x_k}\left(\tau_{kl}^s\right)^n\frac{\partial u_1^n}{\partial x_l}, \frac{\partial \varphi_m^s}{\partial x_1}\right)_{\Omega^{sh}}+\Delta t^2\left(\frac{\partial \varphi_b^s}{\partial x_k}\left(\tau_{kl}^s\right)^n\frac{\partial u_2^n}{\partial x_l},
\frac{\partial \varphi_m^s}{\partial x_2}\right)_{\Omega^{sh}} \\
& +2\Delta t\left(\frac{\partial \varphi_b^s}{\partial x_k}\left(\tau_{k1}^s\right)^n, \frac{\partial \varphi_m^s}{\partial x_1}\right)_{\Omega^{sh}}+\Delta t\left(\frac{\partial \varphi_b^s}{\partial x_k}\left(\tau_{k2}^s\right)^n,
\frac{\partial \varphi_m^s}{\partial x_2}\right)_{\Omega^{sh}}.
\end{split}
\end{equation*}
It can be seen from the pattern of the above matrices that one can get ${\bf K}_{22}^s$ by changing the subscript $1$ to $2$, and changing $2$ to $1$ in the formula of ${\bf K}_{11}^s$. Similarly, the elements of ${\bf K}_{12}^s$ can be expressed as:
\begin{equation*}
\begin{split}
&\left({\bf K}_{12}^s\right)_{bm}=\mu^s \Delta t\left(\frac{\partial \varphi_b^s}{\partial x_1},
\frac{\partial \varphi_m^s}{\partial x_2}\right)_{\Omega^{sh}}
+\mu^s \Delta t^2\left(\frac{\partial u_1^n}{\partial x_k}\frac{\partial \varphi_b^s}{\partial x_k},
\frac{\partial \varphi_m^s}{\partial x_2}\right)_{\Omega^{sh}}\\
&+\Delta t^2\left(\frac{\partial u_1^n}{\partial x_k}\left(\tau_{kl}^s\right)^n\frac{\partial \varphi_b^s}{\partial x_l},
\frac{\partial \varphi_m^s}{\partial x_2}\right)_{\Omega^{sh}}
+\Delta t\left(\left(\tau_{1k}^s\right)^n\frac{\partial \varphi_b^s}{\partial x_k},
\frac{\partial \varphi_m^s}{\partial x_2}\right)_{\Omega^{sh}},
\end{split}
\end{equation*}
and $\left({\bf K}_{21}^s\right)_{bm}=\left({\bf K}_{12}^s\right)_{mb}$, $\left(b,m=1,2,\cdots N^s\right)$. \\
The matrix  $\bf B$ has the following expression.
\begin{equation}
{\bf B}=
\begin{bmatrix}
{\bf B}_1 \\
{\bf B}_2 \\
\end{bmatrix}
,
\end{equation}
where
$$
\left({\bf B}_1\right)_{mk}=\left(\psi_k, \frac{\partial \varphi_m}{\partial x_1}\right)_{\Omega^h},
\left({\bf B}_2\right)_{mk}=\left(\psi_k, \frac{\partial \varphi_m}{\partial x_2}\right)_{\Omega^h}
$$
$\left(k=1,2,\cdots N^p\right. $ and $\left. m=1,2,\cdots N^f\right)$. The vector
\begin{equation}
{\bf f}=
\begin{pmatrix}
{\bf f}_1 \\
{\bf f}_2 \\
\end{pmatrix}
\end{equation}
is the fluid force vector, where
$$
\left({\bf f}_1\right)_m=\rho^f\left(g_1,\varphi_m\right)_{\Omega^h}+\left(\bar{h}_1,\varphi_m\right)_{\Gamma^{Nh}},
$$
and
$$
\left({\bf f}_2\right)_m=\rho^f\left(g_2,\varphi_m\right)_{\Omega^h}+\left(\bar{h}_2,\varphi_m\right)_{\Gamma^{Nh}}
$$
$\left(m=1,2,\cdots N^f\right)$. The vector 
\begin{equation}
{\bf f}^s=
\begin{pmatrix}
{\bf f}_1^s \\
{\bf f}_2^s \\
\end{pmatrix}
\end{equation}
is the solid force vector, where
\begin{equation*}
\begin{split}
& \left({\bf f}_1^s\right)_m=\left(\rho^s-\rho^f\right)\left(g_1,\varphi_m^s\right)_{\Omega^{sh}} \\
&+\left(\mu^s\Delta t^2\frac{\partial u_1^n}{\partial x_k}\frac{\partial u_j^n}{\partial x_k} 
 +\Delta t^2\frac{\partial u_1^n}{\partial x_k}\left(\tau_{kl}^s\right)^n\frac{\partial u_j^n}{\partial x_l}-\left(\tau_{1j}^s\right)^n, \frac{\partial \varphi_m^s}{\partial x_j}
\right)_{\Omega^{sh}}
\end{split}
\end{equation*}
and
\begin{equation*}
\begin{split}
& \left({\bf f}_2^s\right)_m=\left(\rho^s-\rho^f\right)\left(g_2,\varphi_m^s\right)_{\Omega^{sh}} \\
&+\left(\mu^s\Delta t^2\frac{\partial u_2^n}{\partial x_k}\frac{\partial u_j^n}{\partial x_k} 
+\Delta t^2\frac{\partial u_2^n}{\partial x_k}\left(\tau_{kl}^s\right)^n\frac{\partial u_j^n}{\partial x_l}-\left(\tau_{2j}^s\right)^n, \frac{\partial \varphi_m^s}{\partial x_j}
\right)_{\Omega^{sh}}
\end{split}
\end{equation*}
$\left(m=1,2,\cdots N^s\right)$.
Finally, matrix ${\bf D}$ is the FEM interpolation matrix which can be expressed as:
\begin{equation}
{\bf D}=
\begin{bmatrix}
{\bf R}^{\rm T} & {} \\
{} & {\bf R}^{\rm T} \\
\end{bmatrix}
.
\end{equation}

Using the arbitrariness of our test vectors $\widetilde{\bf v}$ and $\widetilde{\bf q}$, one can obtain the following linear algebraic equation for the whole FSI system from equation (\ref{energyformoflinearalgbraic}):
\begin{equation}\label{lastlinearalgbraic}
\begin{bmatrix}
{\bf A} & {\bf B} \\
{\bf B}^{\rm T} &{\bf 0}
\end{bmatrix}
\begin{pmatrix}
\widetilde{\bf u} \\
\widetilde{\bf p}
\end{pmatrix}
=
\begin{pmatrix}
\widetilde{\bf b} \\
{\bf 0}
\end{pmatrix}.
\end{equation}

\subsection{The UFEM algorithm}
Having derived a discrete system of equations we now describe the solution algorithm at each time step.
\begin{enumerate}[(1)]
	\item Given the solid configuration $\left({\bf x}^s\right)^n$ and velocity field $\widetilde{\bf u}^n=\left\{\begin{matrix}
	\left(\widetilde{\bf u}^f\right)^n & in \quad \Omega^f \\
	\left(\widetilde{\bf u}^s\right)^n & in \quad \Omega^s
	\end{matrix}\right.$ at time step $n$.
	\item Discretize the convection equation (\ref{lastweakofconvection}) or (\ref{cbs-last}) and solve it to get an intermediate velocity ${\bf u}^*$.
	\item Compute the interpolation matrix  and solve equation (\ref{lastlinearalgbraic}) using ${\bf u}^*$ and $\left(\widetilde{\bf u}^s\right)^n$ as initial values to get velocity field $\widetilde{\bf u}^{n+1}$.
    \item Compute solid velocity $\left(\widetilde{\bf u}^s\right)^{n+1}={\bf D}\widetilde{\bf u}^{n+1}$ and update the solid mesh by $\left({\bf x}^s\right)^{n+1}=\left({\bf x}^s\right)^n+\Delta t\left(\widetilde{\bf u}^s\right)^{n+1}$, then go to step (1) for the next time step.
\end{enumerate}

{\bf Remark 4} When implementing the UFEM algorithm, it is unnecessary to perform the matrix multiplication ${\bf D}^{\rm T}{\bf K}^s{\bf D}$ in (\ref{initialenergyform}) globally, because the FEM interpolation is locally based. All the matrix operations can be computed based on the local element matrices only. Alternatively, if an iterative solver is used, it is actually unnecessary to compute ${\bf D}^{\rm T}{\bf K}^s{\bf D}$. What an iterative step needs is to compute $\left({\bf D}^{\rm T}{\bf K}^s{\bf D}\right){\bf u}$ for a given vector ${\bf u}$, therefore one can compute ${\bf Du}$ first, then ${\bf K}^s\left({\bf D}{\bf u}\right)$, and last ${\bf D}^{\rm T}\left({\bf K}^s{\bf D}{\bf u}\right)$.

\section{Numerical experiments}

In this section, we present some numerical examples that have been selected to allow us to assess our proposed UFEM. We shall demonstrate the convergence of UFEM in time and space, and compare results obtained by the UFEM with those obtained using monolithic approaches and IFEM, as well as compare against results from laboratory experiment. 

In order to improve the computational efficiency, an adaptive spatial mesh with hanging nodes is used in all the following numerical experiments. Readers can reference Appendix A for details of the treatment of hanging nodes.

\subsection{Oscillation of a flexible leaflet oriented across the flow direction}
\begin{figure}[h!]
	\centering
	\includegraphics[width=4in,angle=0]{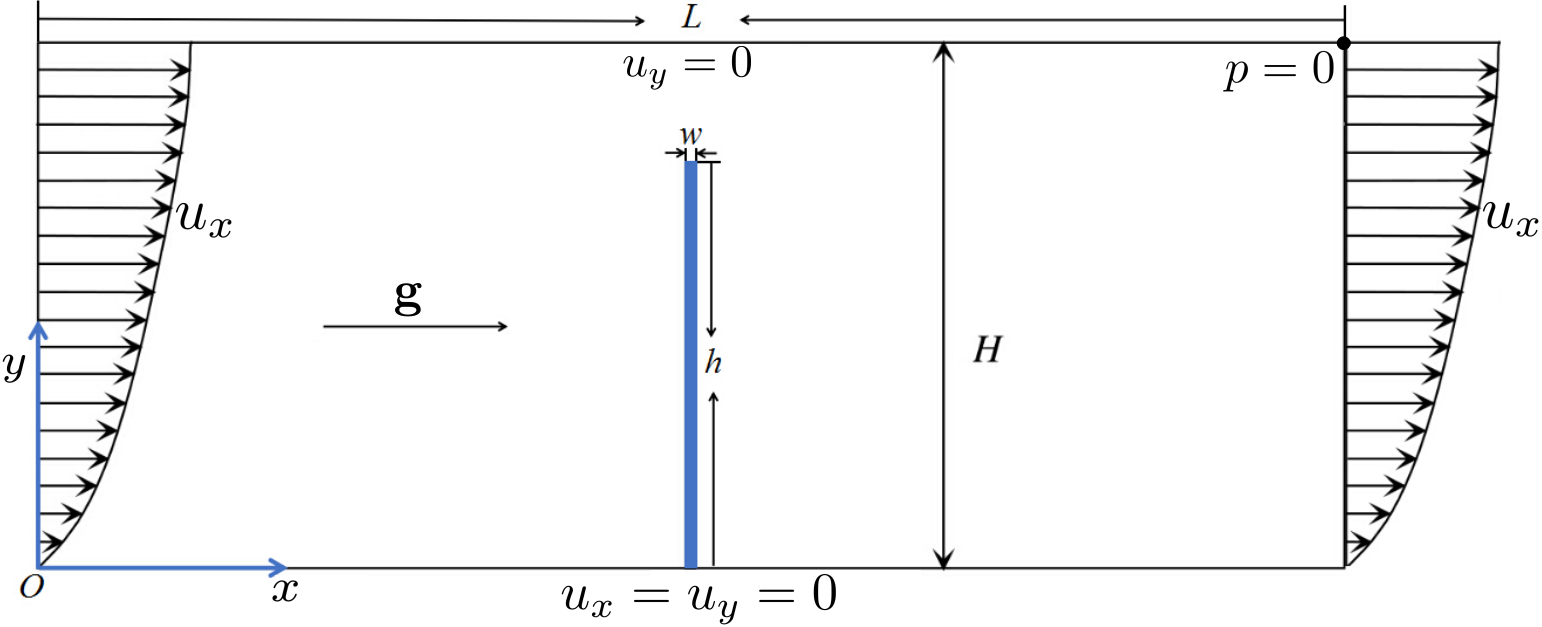}
	\captionsetup{justification=centering}
	\caption {\scriptsize Computational domain and boundary conditions, taken from \cite{baaijens2001fictitious}.} 
	\label{fig2}
\end{figure}

This numerical example is used by \cite{Yu_2005,baaijens2001fictitious,Kadapa_2016} to validate their methods. We first use the same parameters as used in the above three publications in order to compare results and test convergence in time and space, then use a range of parameters to show the robustness of our UFEM. The implicit Least-squares method is used to treat the convection step in all these tests unless otherwise stated. The computational domain and boundary conditions are illustrated in Figure \ref{fig2}.

The inlet flow is in the x-direction and given by $u_x=15.0y\left(2-y\right)sin\left(2\pi t\right)$. Gravity is not considered in the first test (i.e. ${\bf g}=0$), and other fluid and solid properties are presented in Table \ref{Properties and domain size for test problem with a leaflet in a channel}.
\begin{table}[h!]
	\centering
	\begin{tabular}{|c|c|}
		\hline
		Fluid  & Leaflet \\
		\hline
		$L=4.0$ $m$ & $w=0.0212$ $m$ \\
		$H=1.0$ $m$ & $h=0.8$ $m$ \\
		$\rho^f=100$ $\left.kg\right/{m^3}$ & $\rho^s=100$ $\left.kg\right/{m^3}$ \\
		$\mu^f=10$ $\left.N\cdot s\right/{m^2}$ & $\mu^s=10^7$ $\left.N\right/{m^2}$ \\
		\hline
	\end{tabular}
	\captionsetup{justification=centering}
	\caption{Properties and domain size for test problem \\ with a leaflet oriented across the flow direction.}
	\label{Properties and domain size for test problem with a leaflet in a channel}
\end{table}

\begin{figure}[h!]
\centering
    \subfigure[h][$t=0.1s$]{ 
	\includegraphics[width=4.7in,angle=0]{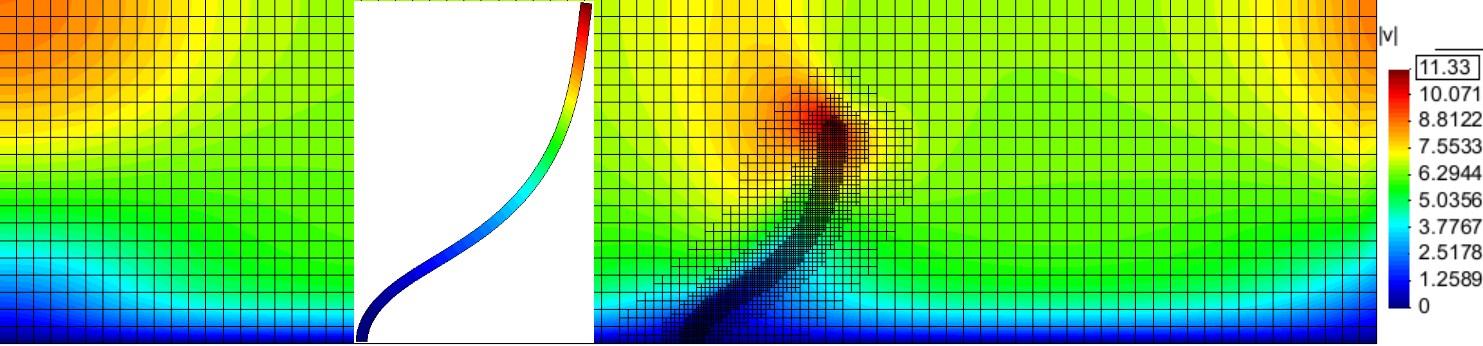}
    }
    \subfigure[$t=0.2s$]{ 
    \includegraphics[width=4.7in,angle=0]{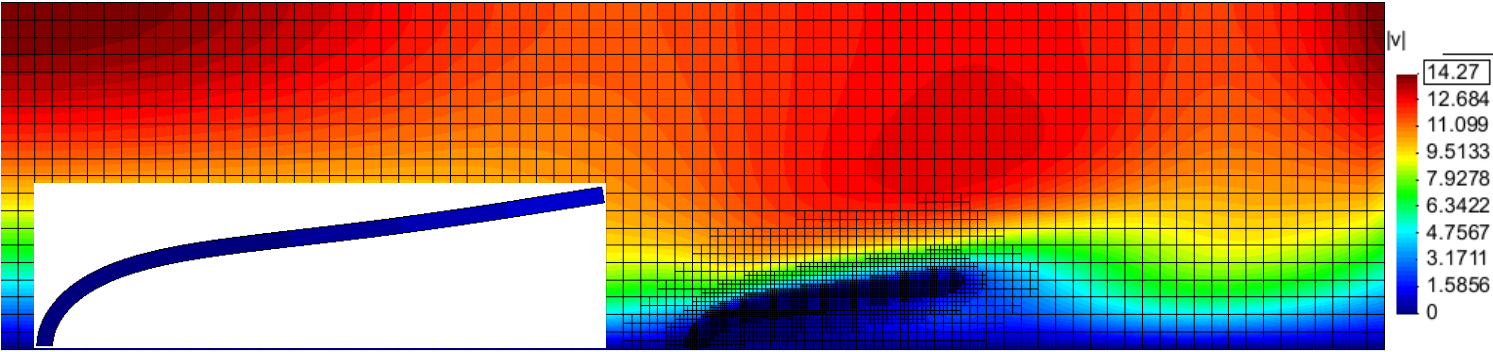}
    }    
    \subfigure[$t=0.6s$]{ 
    	\includegraphics[width=4.7in,angle=0]{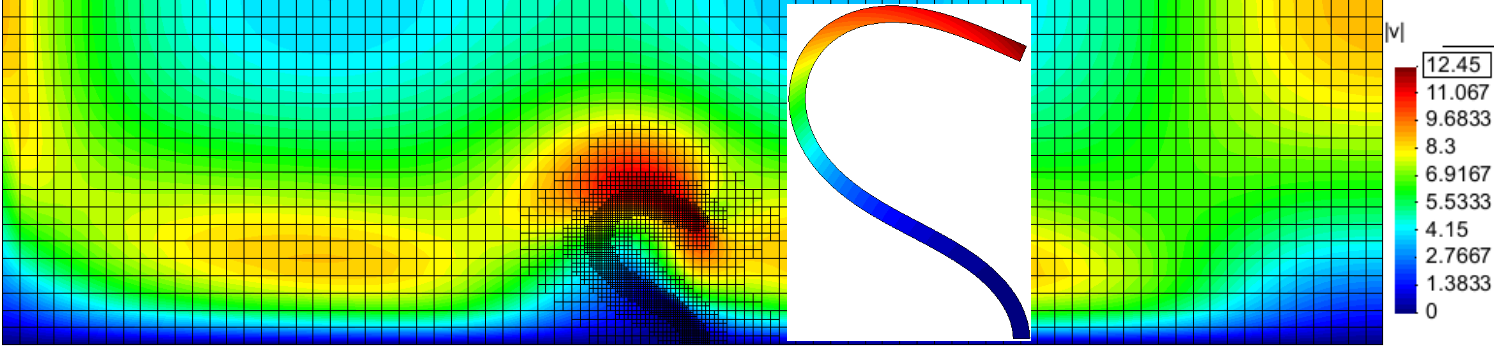}
    }        
    \subfigure[$t=0.8s$]{ 
    	\includegraphics[width=4.7in,angle=0]{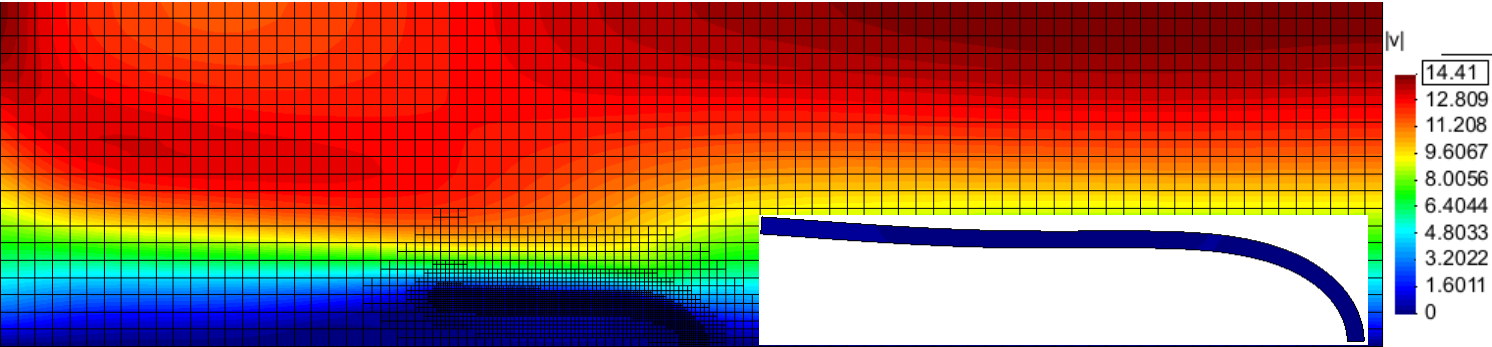}
    }     
\captionsetup{justification=centering}
\caption {\scriptsize Configuration of leaflet and magnitude of velocity on the adaptive fluid mesh.} 
\label{Configuration of leaflet and magnitude of velocity on the adaptive fluid mesh}
\end{figure}

The leaflet is approximated with 1200 linear triangles with 794 nodes (medium mesh size), and the corresponding fluid mesh is adaptive in the vicinity of the leaflet so that it has a similar size. A stable time step $\Delta t=5.0\times 10^{-4}s$ is used in these initial simulations. The configuration of the leaflet is illustrated at different times in Figure \ref{Configuration of leaflet and magnitude of velocity on the adaptive fluid mesh}.

\begin{figure}[h!]
	\begin{minipage}[t]{0.5\linewidth}
    \centering  
    \includegraphics[width=2.5in,angle=0]{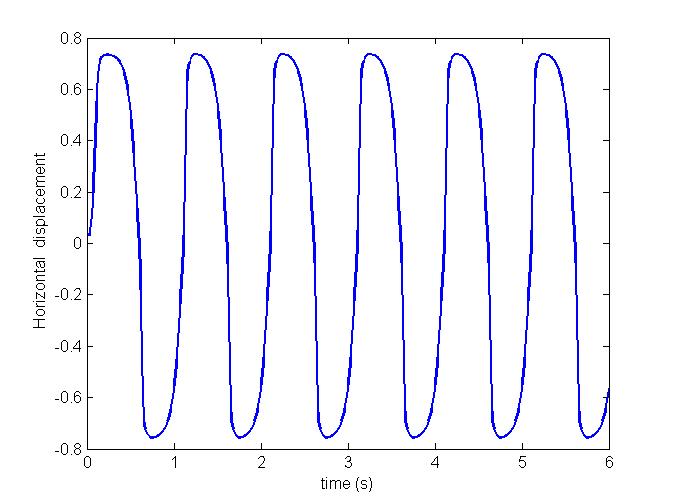}
    \end{minipage}
	\begin{minipage}[t]{0.5\linewidth}
		\centering  
		\includegraphics[width=2.5in,angle=0]{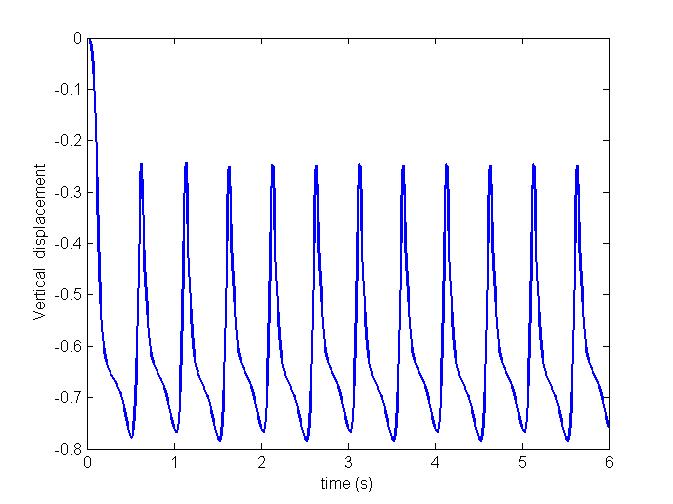}
	\end{minipage}   
	\captionsetup{justification=centering}
	\caption {\scriptsize Evolution of horizontal and vertical displacement at top right corner of the leaflet.} 
	\label{Evolution of horizontal and vertical displacement at top right corner of the leaflet}
\end{figure}

Previously published numerical results are qualitatively similar to those in Figure \ref{Configuration of leaflet and magnitude of velocity on the adaptive fluid mesh} but show some quantitative variations. For example, \cite{baaijens2001fictitious} solved a fully-coupled system but the coupling is limited to a line, and the solid in their results (Figure 7 (l)) behaves as if it is slightly harder. Alternatively, \cite{Yu_2005} used a fractional step scheme to solve the FSI equations combined with a penalty method to enforce the incompressibility condition. In their results (Fig. 3 (h)) the leaflet behaves as if it is slightly softer than \cite{baaijens2001fictitious} and harder than \cite{Kadapa_2016}. In \cite{Kadapa_2016} a beam formulation is used to describe the solid. The fluid mesh is locally refined using hierarchical B-Splines, and the FSI equation is solved monolithically. The leaflet in their results (Fig. 34) behaves as softer than the other two considered here. Our results in Figure \ref{Configuration of leaflet and magnitude of velocity on the adaptive fluid mesh} are most similar to those of \cite{Kadapa_2016}. This may be seen more precisely by inspection of the graphs of the oscillatory motion of the leaflet tip in Figure \ref{Evolution of horizontal and vertical displacement at top right corner of the leaflet} corresponding to Fig. 32 in \cite{Kadapa_2016}. We point out here that the explicit Taylor-Galerkin method is also used to solve the convection step for this test, and we gain almost the same accuracy using the same time step $\Delta t=5.0\times 10^{-4}s$. 

Having validated our results for this example against the work of others, we shall use this test case to further explore more details of our method. 

\begin{figure}[h!]
	\centering
	\includegraphics[width=3.5in,angle=0]{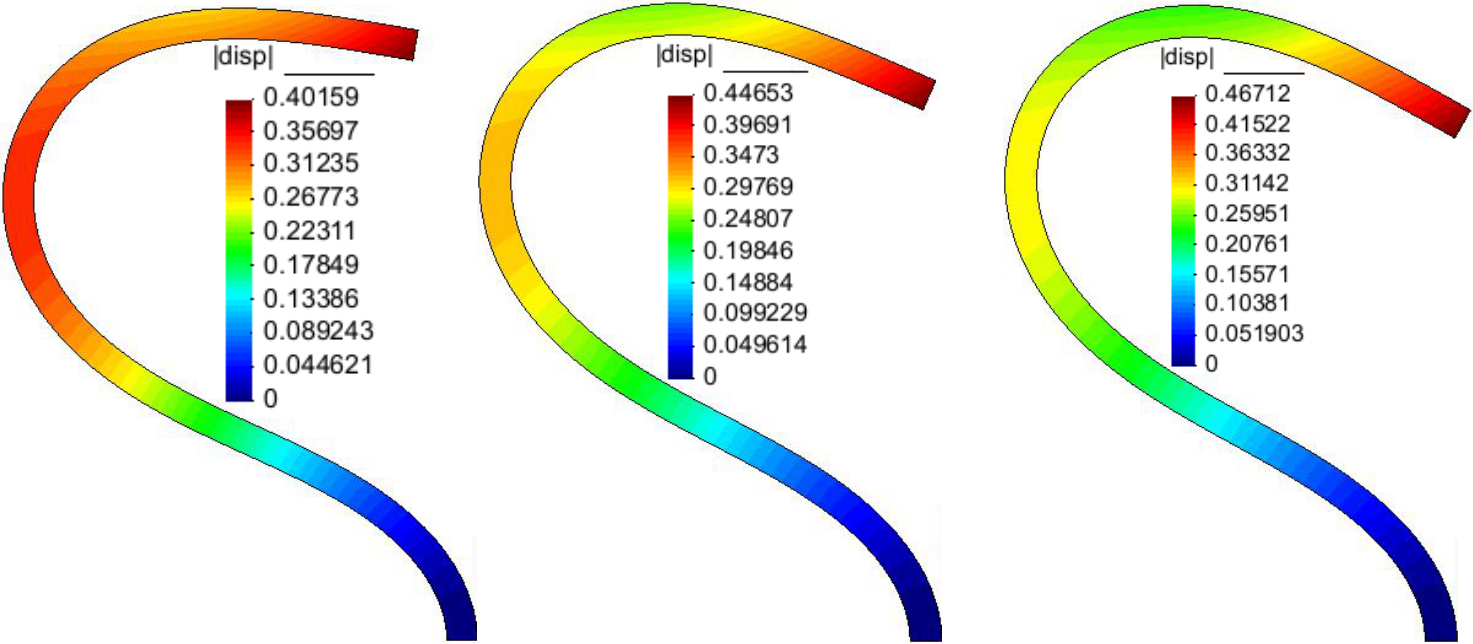}
	\captionsetup{justification=centering}
	\caption*{\scriptsize (a)coarse \qquad\qquad\qquad\qquad  (b)medium \qquad\qquad\qquad\qquad (c)fine}	
	\caption {\scriptsize Configuration of leaflet for different mesh ratio $r_m$, \\
		and contour plots of displacement magnitude at $t=0.6s$.} 
	\label{mesh ratio test}
\end{figure}

We commence by testing the influence of the ratio of fluid and solid mesh sizes $r_m$=(local fluid element area)/(solid element area). Fixing the fluid mesh size, three different solid mesh sizes are chosen: coarse (640 linear triangles with 403 nodes $r_m\approx1.5$), medium (1200 linear triangles with 794 nodes $r_m\approx3.0$) and fine (2560 linear triangles with 1445 nodes $r_m\approx5.0$), and a stable time step $\Delta t=5.0\times 10^{-4}s$ is used. From these tests we observe that there is a slight difference in the solid configuration for different meshes, as illustrated at $t=0.6s$ in Figure \ref*{mesh ratio test}, however the difference in displacement decreases as the solid mesh becomes finer. Further, we found that $1.5\leq r_m\leq5.0$ ensures the stability of the proposed UFEM approach. Note that we use a 9-node quadrilateral for the fluid velocity and 3-node triangle for solid velocity, so $r_m\approx3.0$ means the fluid and solid mesh locally have a similar number of nodes for velocity.

\begin{figure}[h!]
	\begin{minipage}[t]{0.5\linewidth}
		\centering  
		\includegraphics[width=2in,angle=0]{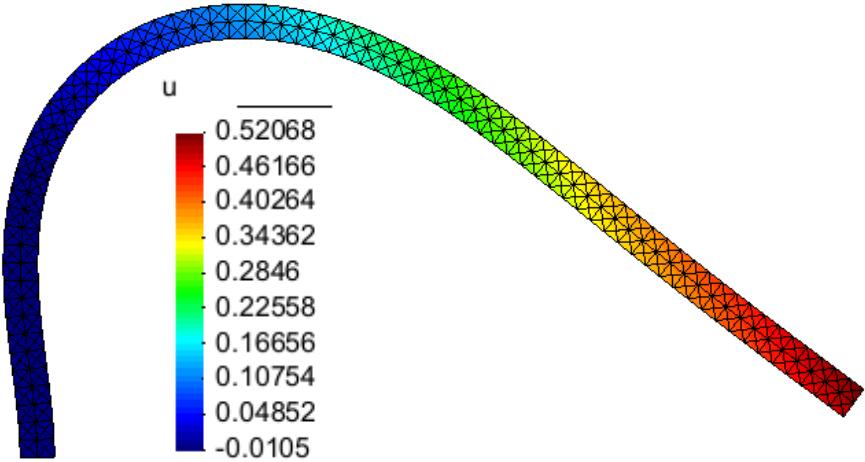}	
		\caption*{\scriptsize(a) Coarse}
	\end{minipage}
	\begin{minipage}[t]{0.5\linewidth}
		\centering  
		\includegraphics[width=2in,angle=0]{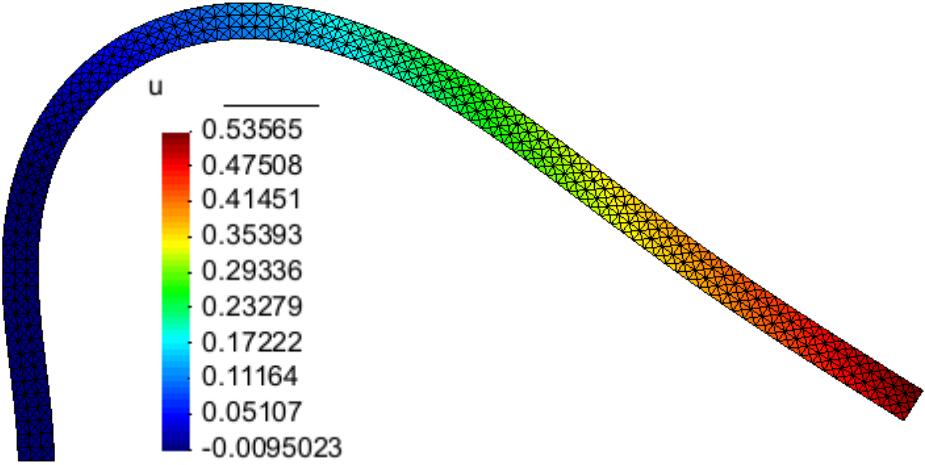}
		\caption*{\scriptsize(b) Medium}
	\end{minipage}   
	\begin{minipage}[t]{0.5\linewidth}
		\centering  
		\includegraphics[width=2in,angle=0]{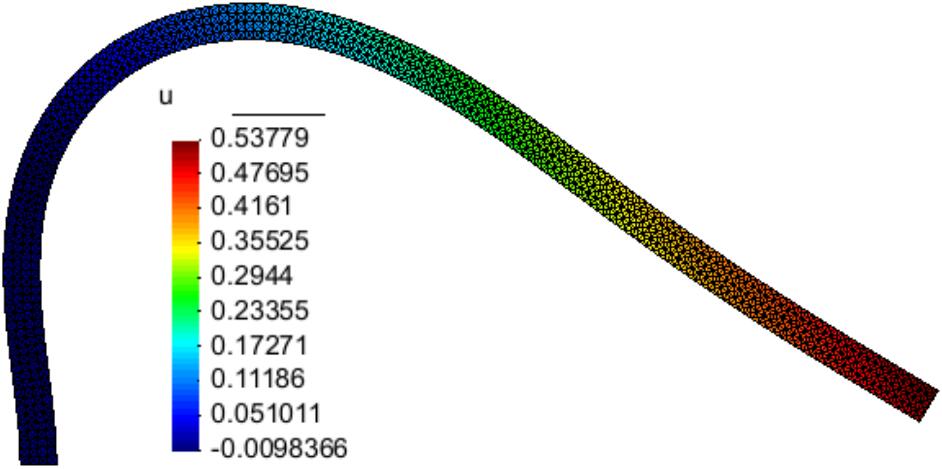}
        \caption*{\scriptsize(c) Fine}
	\end{minipage}
	\begin{minipage}[t]{0.5\linewidth}
		\centering  
		\includegraphics[width=2in,angle=0]{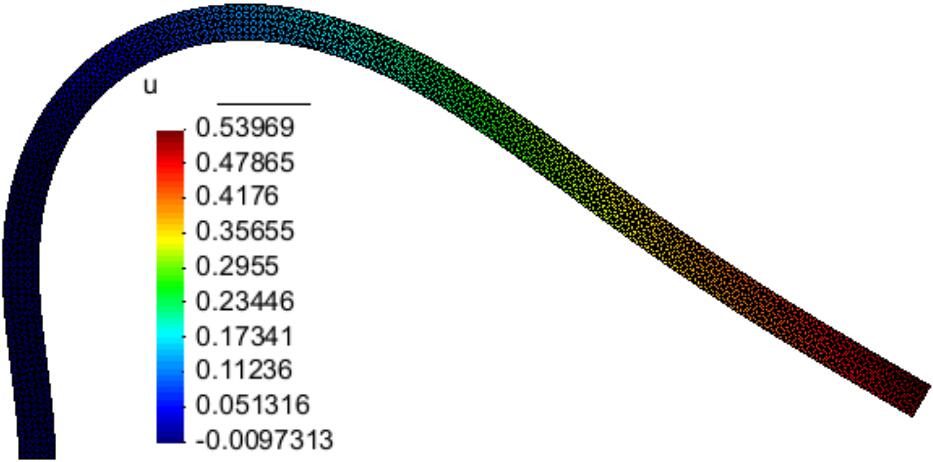}
		\caption*{\scriptsize(d) Very fine}
	\end{minipage}   		
	\captionsetup{justification=centering}
	\caption {\scriptsize Contour plots of horizontal velocity at $t=0.5s$.} 
	\label{mesh convergence test}
\end{figure}

We next consider convergence tests undertaken for refinement of both the fluid and solid meshes with the fixed ratio of mesh sizes $r_m\approx3.0$. Four different levels of meshes are used, the solid meshes are: coarse (584 linear triangles with 386 nodes), medium (1200 linear triangles with 794 nodes), fine (2560 linear triangles with 1445 nodes), and very fine (3780 linear triangles with 2085 nodes). The fluid meshes have the corresponding sizes with the solid at their maximum refinement level. As can be seen in Figure \ref{mesh convergence test} and Table \ref{Comparison of maximum velocity for different meshes}, the velocity is converging as the mesh becomes finer.

\begin{table}[h!]
\newcommand{\tabincell}[2]{\begin{tabular}{@{}#1@{}}#2\end{tabular}}	
	\centering
	\begin{tabular}{|c|c|}
		\hline
		{Between different mesh sizes}  & \tabincell{c} {Difference of maximum \\  horizontal velocity at $t=0.5s$} \\
		\hline
		coarse and medium & 0.01497\\
		\hline
		medium and fine & 0.00214 \\
		\hline
		fine and very fine & 0.00190 \\
		\hline
	\end{tabular}
	\caption{Comparison of maximum velocity for different meshes.}
	\label{Comparison of maximum velocity for different meshes}
\end{table}

\begin{figure}[h!]
	\begin{minipage}[t]{0.5\linewidth}
		\centering  
		\includegraphics[width=2in,angle=0]{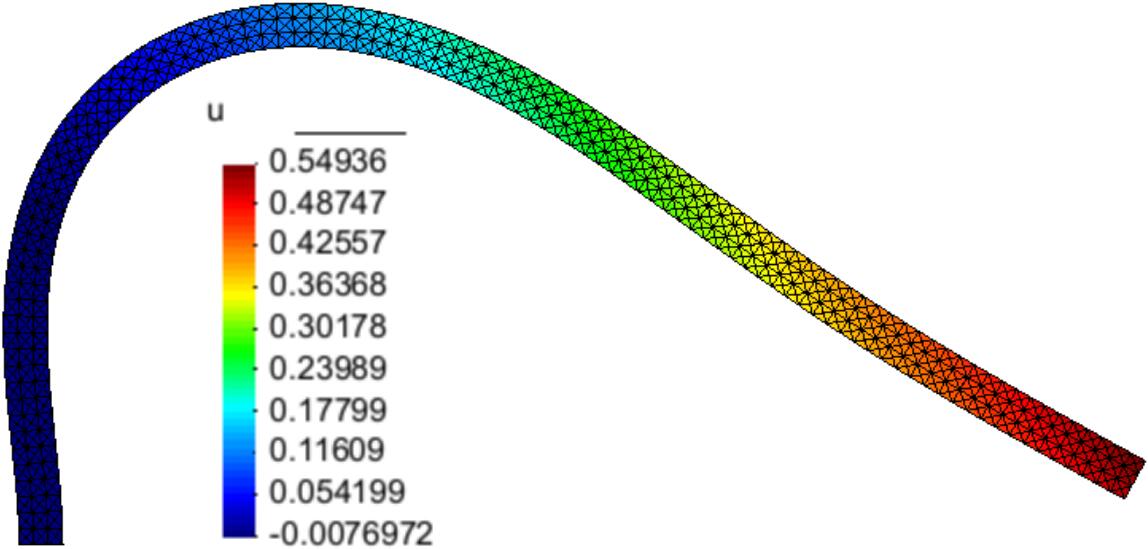}	
        \captionsetup{justification=centering}
		\caption*{\scriptsize(a) $\Delta t=2.0\times 10^{-3}s$ \\ (breaks down at $t=0.61s$).}
	\end{minipage}
	\begin{minipage}[t]{0.5\linewidth}
		\centering  
		\includegraphics[width=2in,angle=0]{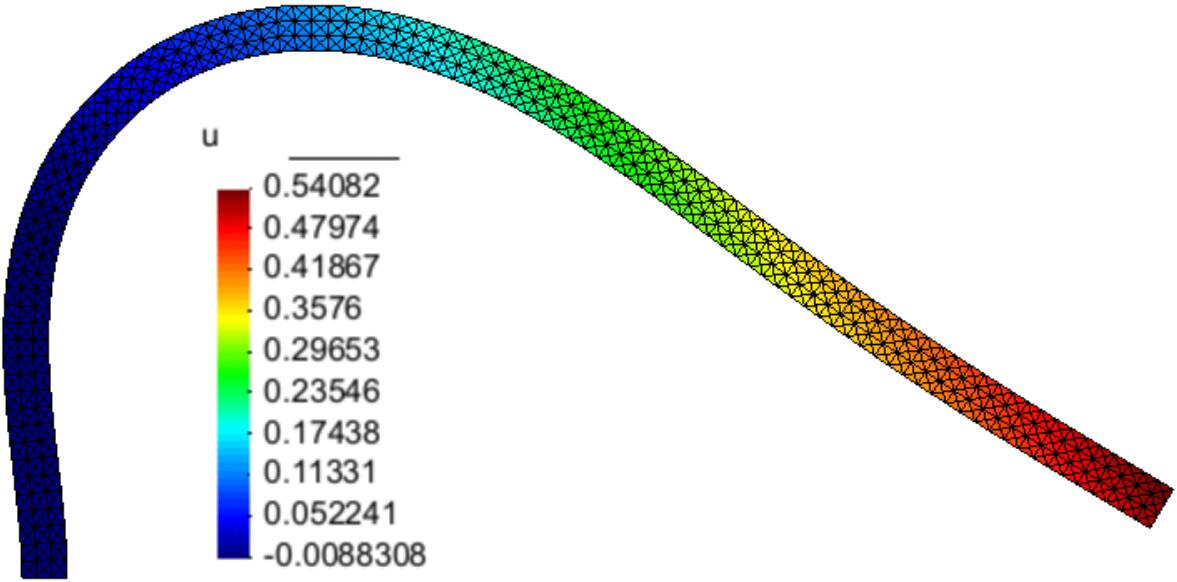}
		\caption*{\scriptsize(b) $\Delta t=1.0\times 10^{-3}s$.}
	\end{minipage}   
	\begin{minipage}[t]{0.5\linewidth}
		\centering  
		\includegraphics[width=2in,angle=0]{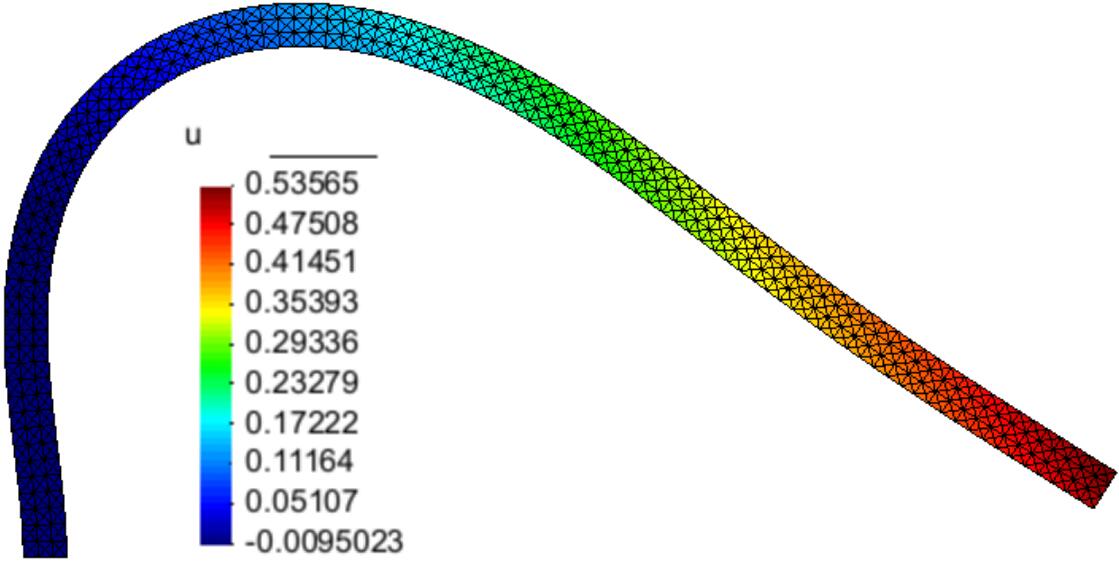}
		\caption*{\scriptsize(c) $\Delta t=5.0\times 10^{-4}s$.}
	\end{minipage}
	\begin{minipage}[t]{0.5\linewidth}
		\centering  
		\includegraphics[width=2in,angle=0]{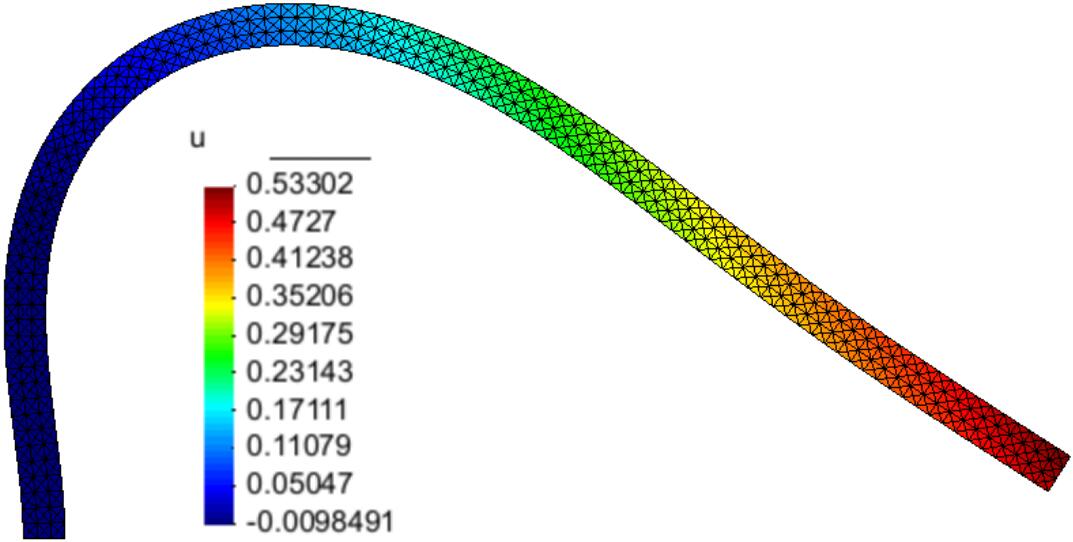}
		\caption*{\scriptsize(d) $\Delta t=2.5\times 10^{-4}s$.}
	\end{minipage}   		
	\captionsetup{justification=centering}
	\caption {\scriptsize Contour plots of horizontal velocity at $t=0.5s$.} 
	\label{time convergence test}
\end{figure}

In addition, we consider tests of convergence in time using a fixed ratio of fluid and solid mesh sizes $r_m\approx3.0$. Using the medium solid mesh size and the same fluid mesh size as above, results are shown in Figure \ref{time convergence test} and Table \ref{Comparison of maximum velocity for different time step size}. From these it can be seen that the velocities are converging as the time step decreases.

\begin{table}[h!]
	\newcommand{\tabincell}[2]{\begin{tabular}{@{}#1@{}}#2\end{tabular}}	
	\centering
	\begin{tabular}{|c|c|}
		\hline
		{Steps sizes compared}  & \tabincell{c} {Difference of maximum \\  horizontal velocity at $t=0.5s$} \\
		\hline
		$\Delta t=2.0\times 10^{-3}$ and $\Delta t=1.0\times 10^{-3}$ & 0.00854\\
		\hline
		$\Delta t=1.0\times 10^{-3}$ and $\Delta t=5.0\times 10^{-4}$ & 0.00517 \\
		\hline
		$\Delta t=5.0\times 10^{-4}$ and $\Delta t=2.5\times 10^{-4}$ & 0.00263 \\
		\hline
	\end{tabular}
	\caption{Comparison of maximum velocity for different time step size.}
	\label{Comparison of maximum velocity for different time step size}
\end{table}

\begin{figure}[h!]
	\begin{minipage}[t]{0.5\linewidth}
		\centering  
		\includegraphics[width=1.5in,angle=0]{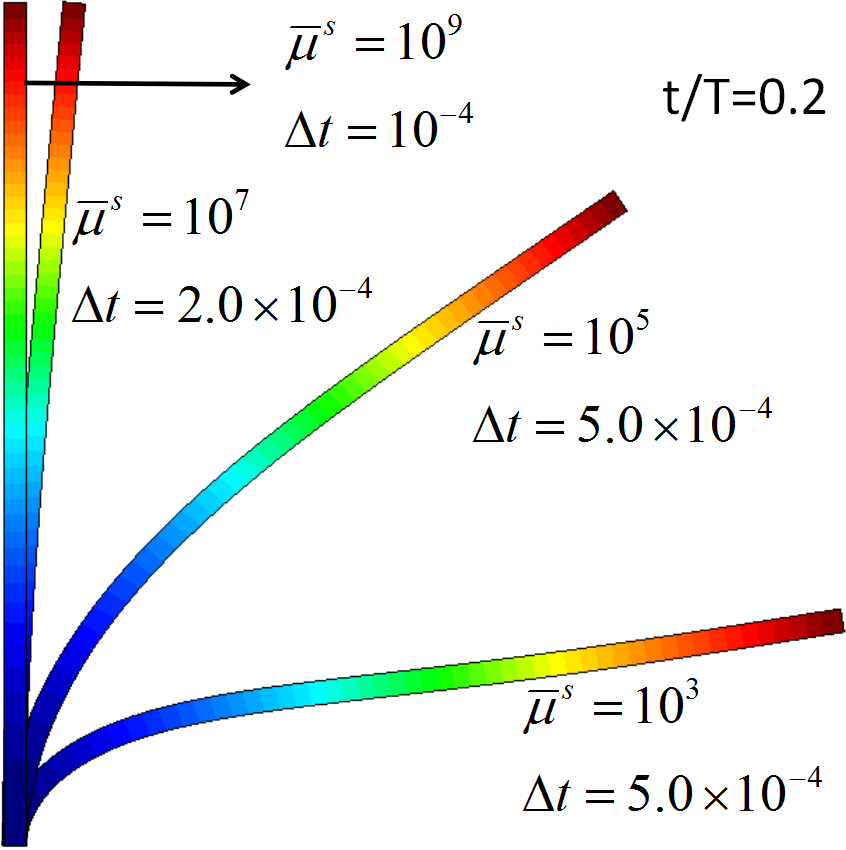}	
		\captionsetup{justification=centering}
		\caption*{\scriptsize(a) $\rho^r=1, Re=100$ and $Fr=0$.}
	\end{minipage}
	\begin{minipage}[t]{0.5\linewidth}
		\centering  
		\includegraphics[width=2.5in,angle=0]{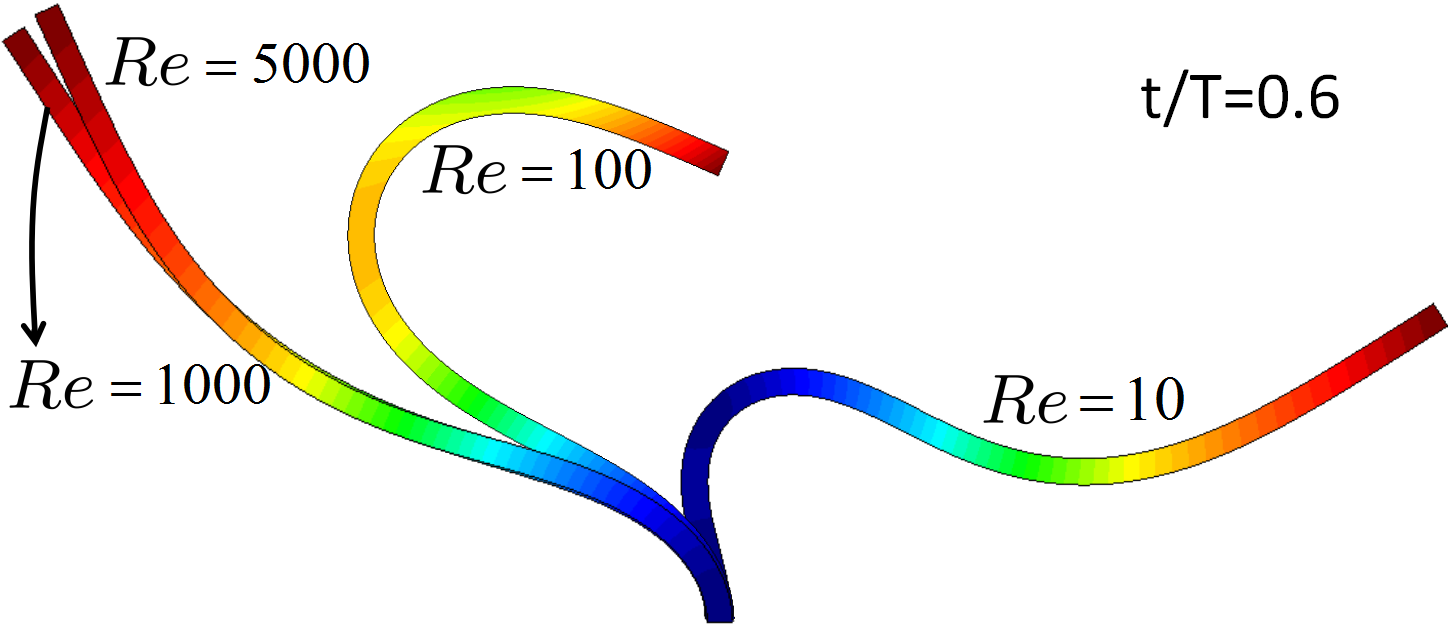}
		\caption*{\scriptsize(b) $\rho^r=1, \bar{\mu}^s=10^3$ and $Fr=0$.}
	\end{minipage}   
	\begin{minipage}[t]{0.5\linewidth}
		\centering  
		\includegraphics[width=1.5in,angle=0]{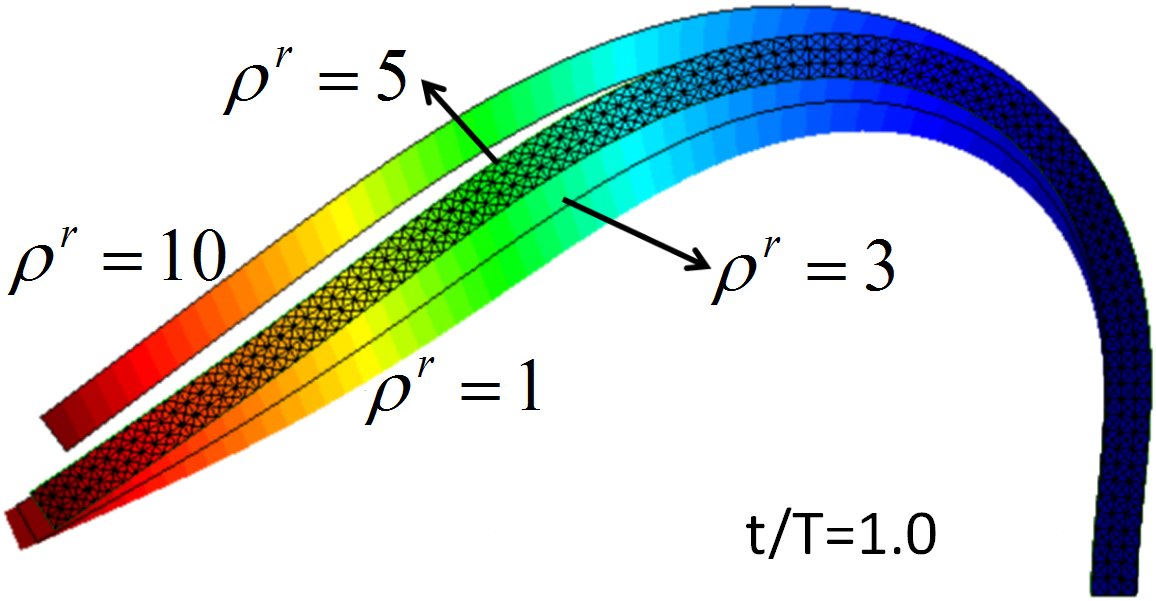}
		\caption*{\scriptsize(c) $Re=100, \bar{\mu}^s=10^3$ and $Fr=0$.}
	\end{minipage}
	\begin{minipage}[t]{0.5\linewidth}
		\centering  
		\includegraphics[width=1.5in,angle=0]{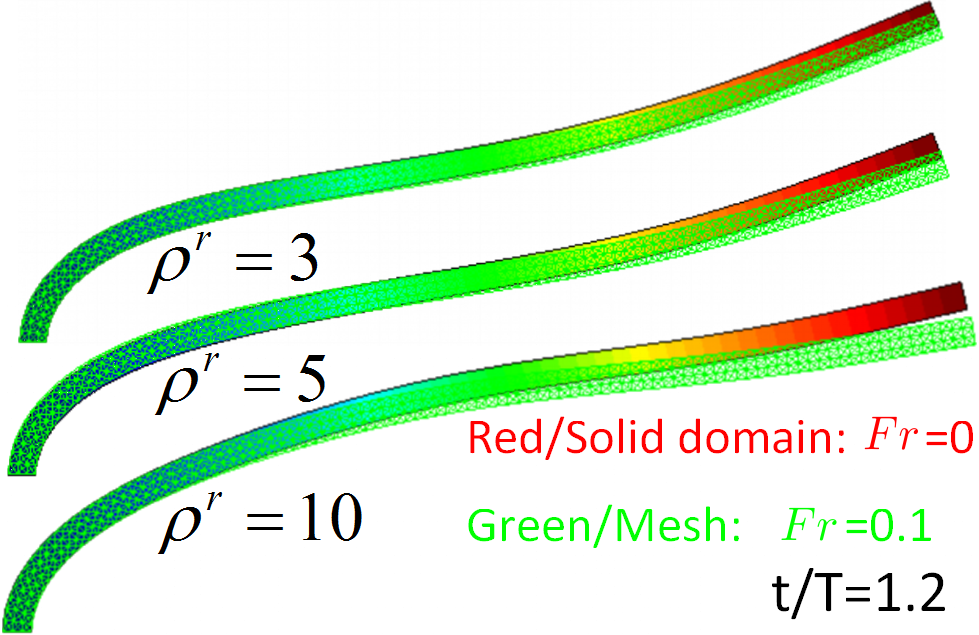}
		\caption*{\scriptsize(d) $Re=100$ and $\bar{\mu}^s=10^3$.}
	\end{minipage}   		
	\captionsetup{justification=centering}
	\caption {\scriptsize Parameters sets and results, $\Delta t=5.0\times 10^{-4}s$ for Group (b)$\sim$(d).} 
	\label{permutation of parameters}
\end{figure}

Finally, in order to assess the robustness of our approach, we vary each of the physical parameters using three different cases as shown in Figure \ref{permutation of parameters}. A medium mesh size with fixed $r_m\approx3.0$ is used to undertake all of these tests. The dimensionless parameters shown in Figure \ref{permutation of parameters} are defined as: $\rho^r=\frac{\rho^s}{\rho^f}, \bar{\mu}^s=\frac{\mu^s}{\rho^fU^2}, Re=\frac{\rho^fUH}{\mu^f}$ and $Fr=\frac{gH}{U^2}$ where the average velocity $U=10$ in this example. $T=1$ is the period of inlet flow.

It can be seen from the results of group (a) that the larger the value of shear modulus $\bar{\mu}^s$ the harder the solid behaves, however a smaller time step is required. For the case of $\bar{\mu}^s=10^9$, the solid behaves almost like a rigid body, as we would expect. From results of group (b), it is clear that the Reynolds Number $\left(Re\right)$ has a large influence on the behavior of the solid. The density and gravity have relatively less influence on the behavior of solid in this problem which can be seen from the results of group (c) and group (d).

\subsection{Oscillation of a flexible leaflet oriented along the flow direction}
\begin{figure}[h!]
	\centering  
	\includegraphics[width=3in,angle=0]{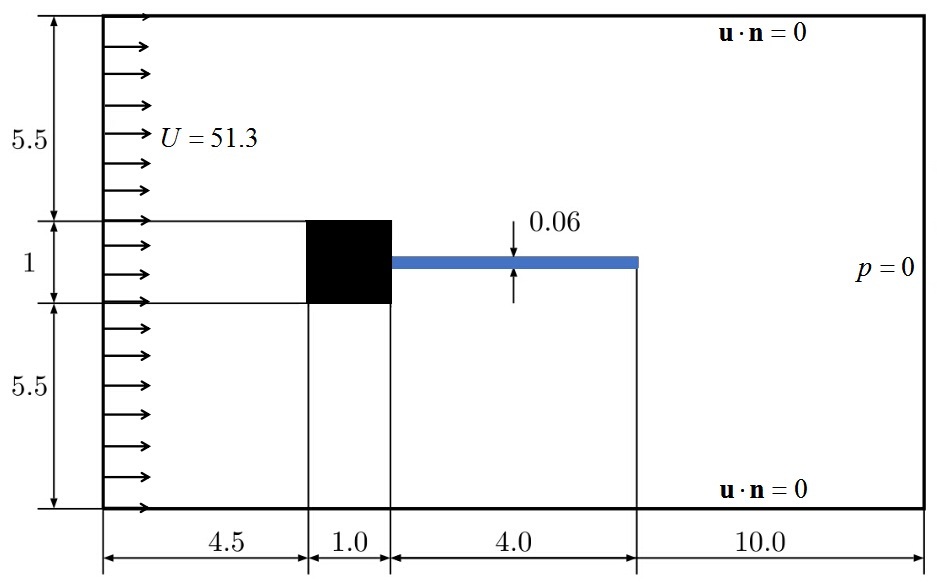}
	\captionsetup{justification=centering}
	\caption {\scriptsize Computational domain and boundary condition for oscillation of flexible leaflet.} 
	\label{Computational domain and boundary condition for oscillation of flexible leaflet}						
\end{figure}

The following test problem that we consider is taken from \cite{wall1999fluid}, which describes an implementation on a ALE fitted mesh. It has since been used as a benchmark to validate different numerical schemes \cite{Kadapa_2016,Hesch_2014}. The geometry and boundary conditions are shown in Figure \ref{Computational domain and boundary condition for oscillation of flexible leaflet}.

\begin{figure}[h!]
	\begin{minipage}[t]{0.5\linewidth}
		\centering  
		\includegraphics[width=2.3in,angle=0]{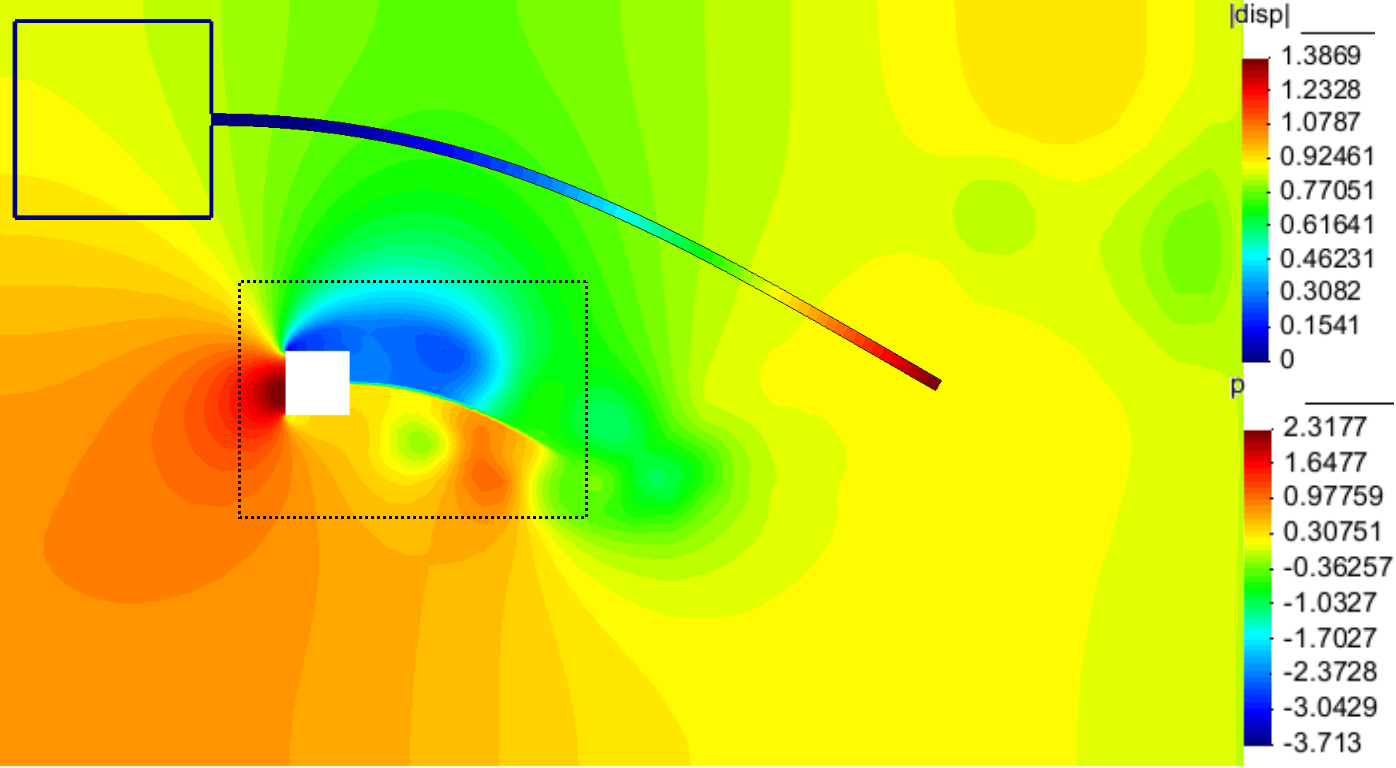}	
        \caption*{\scriptsize(a) Leaflet displacement and fluid pressure.}
	\end{minipage}
	\begin{minipage}[t]{0.5\linewidth}
		\centering  
		\includegraphics[width=2.3in,angle=0]{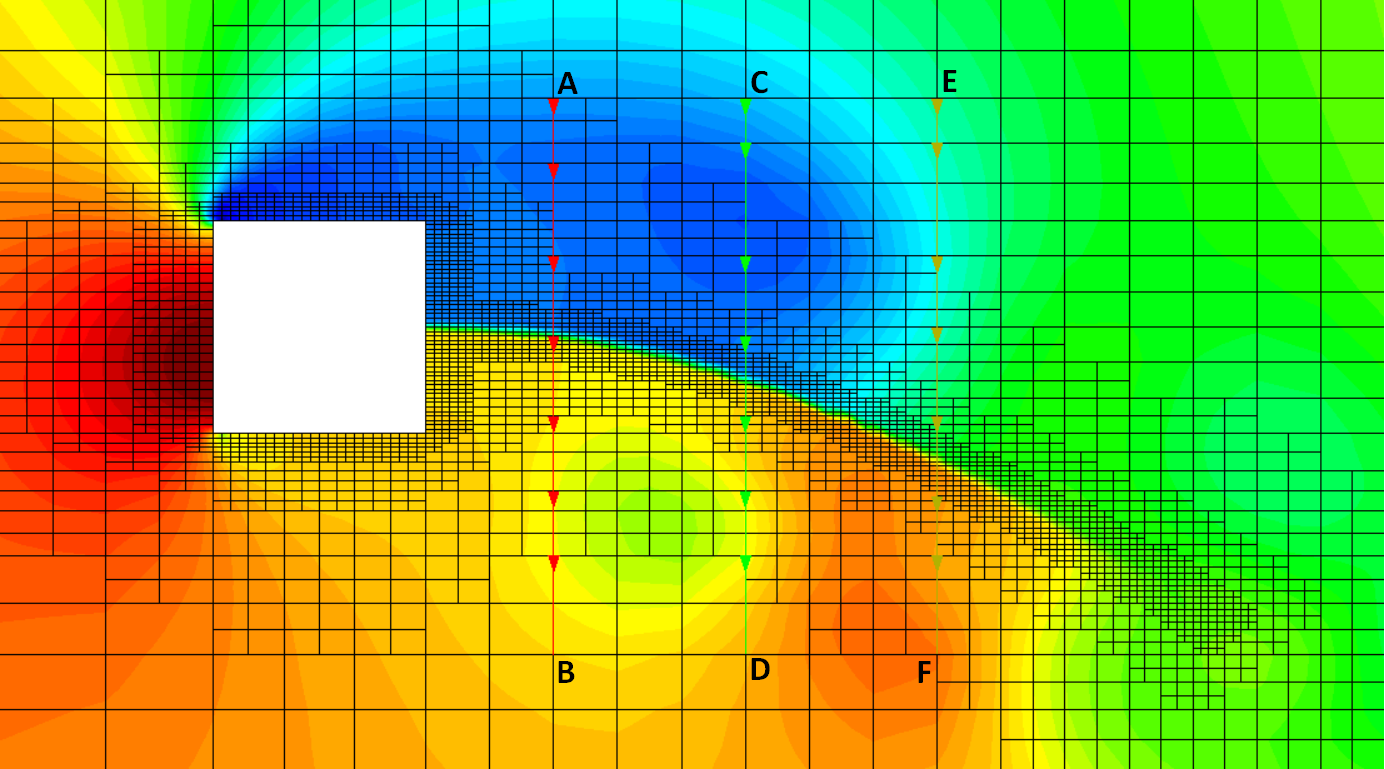}
        \caption*{\scriptsize(b) Mesh refinement near the structure.}		
	\end{minipage}     		
	\captionsetup{justification=centering}
	\caption {\scriptsize Contour plots of leaflet displacement and fluid pressure.} 
	\label{Contours of leaflet displacement and fluid pressure}	
\end{figure}

\begin{figure}[h!]
	\centering  
	\includegraphics[width=3.5in,angle=0]{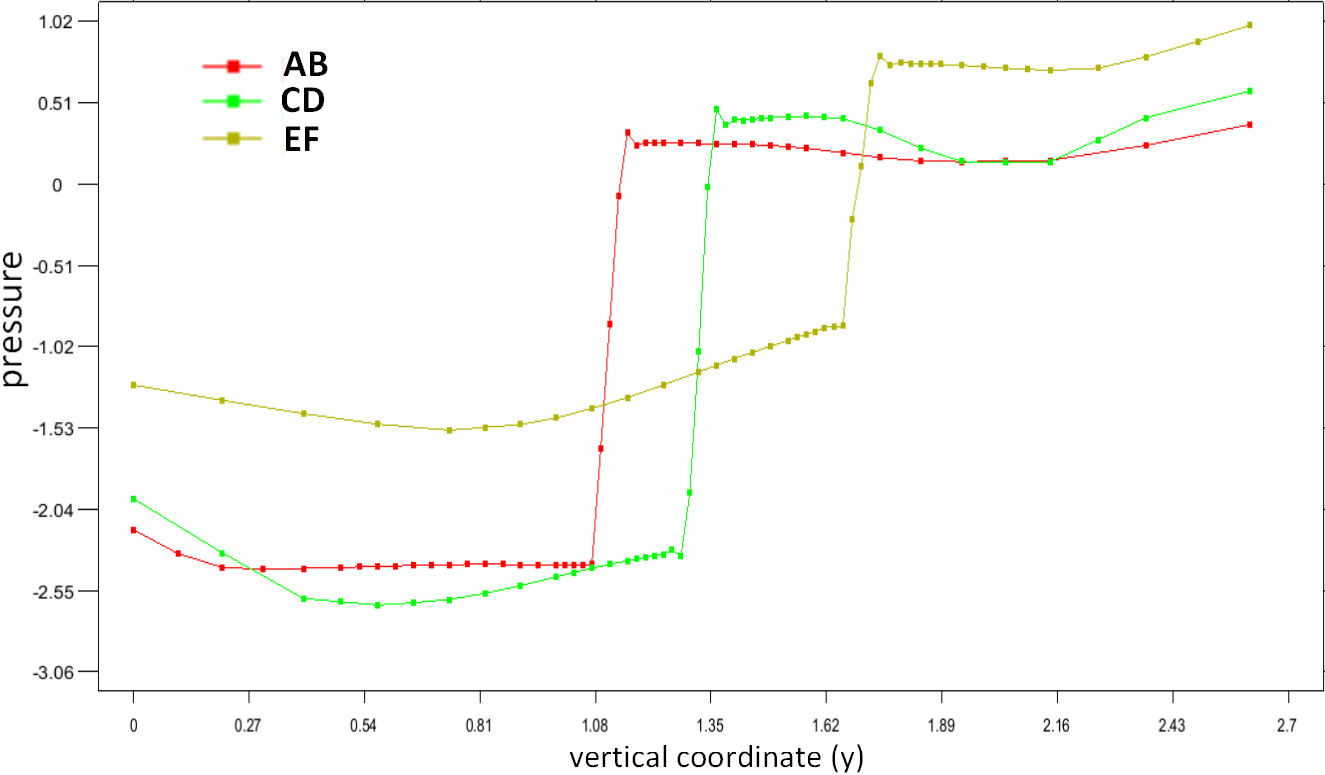}
	\captionsetup{justification=centering}
	\caption {\scriptsize Distribution of pressure across the leaflet on the three lines in Figure \ref{Contours of leaflet displacement and fluid pressure} (b).} 
	\label{Distribution of pressure across the leaflet}						
\end{figure}

For the fluid, the viscosity and density are $\mu^f=1.82\times 10^{-4}$ and $\rho^f=1.18\times 10^{-3}$ respectively. For the solid, we use shear modulus $\mu^s=9.2593\times 10^{-5}$ and density $\rho^s=0.1$. The leaflet is divided by 1063 3-node linear triangles with 666 nodes, and the corresponding fluid mesh locally has a similar node density to the leaflet ($r_m\approx3.0$). First the Least-squares method is tested and a stable time step $\Delta t=1.0\times 10^{-3}s$ is used. A snapshot of the leaflet deformation and fluid pressure at $t=5.44s$ are illustrated in Figure \ref{Contours of leaflet displacement and fluid pressure}. In Figure \ref{Distribution of pressure across the leaflet}, the distributions of pressure across the leaflet corresponding to the three lines (AB, CD and EF) in Figure \ref{Contours of leaflet displacement and fluid pressure} (b) are plotted, from which we can observe that the sharp jumps of pressure across the leaflet are captured.

\begin{figure}[h!]
	\begin{minipage}[t]{0.5\linewidth}
		\centering  
		\includegraphics[width=2in,angle=0]{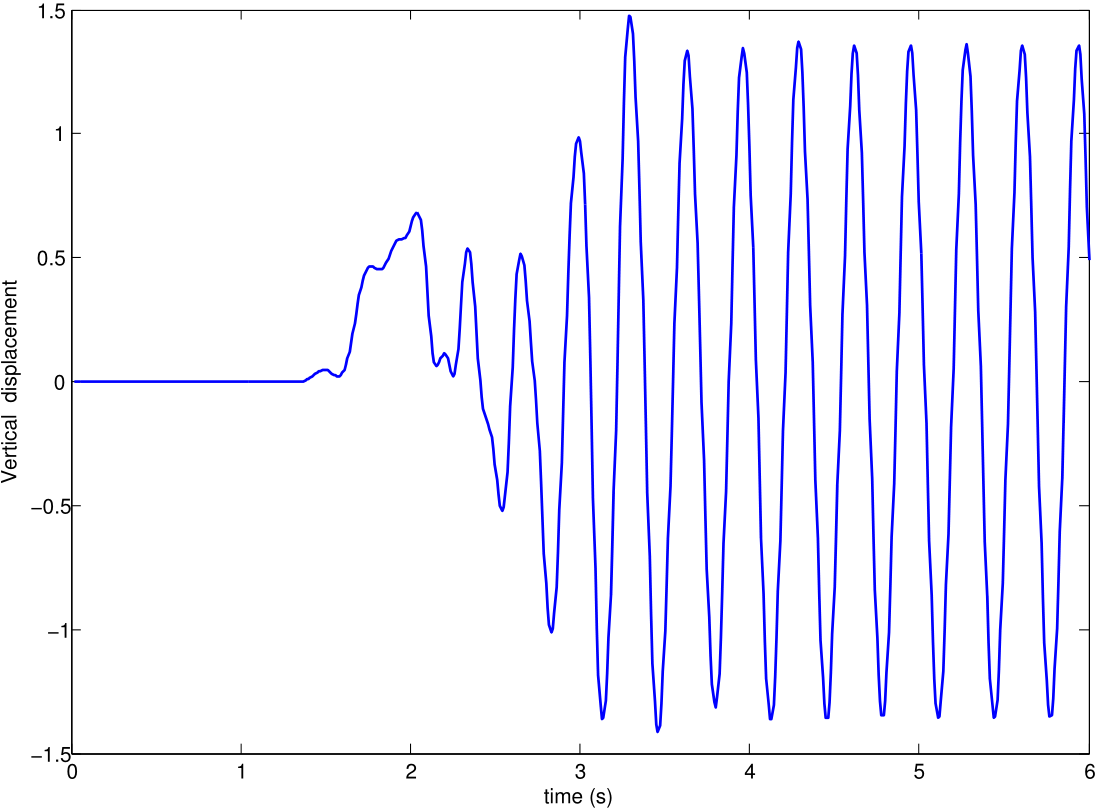}	
		\caption*{\scriptsize (a) Implicit Least-squares method.}
	\end{minipage}
	\begin{minipage}[t]{0.5\linewidth}
		\centering  
		\includegraphics[width=2in,angle=0]{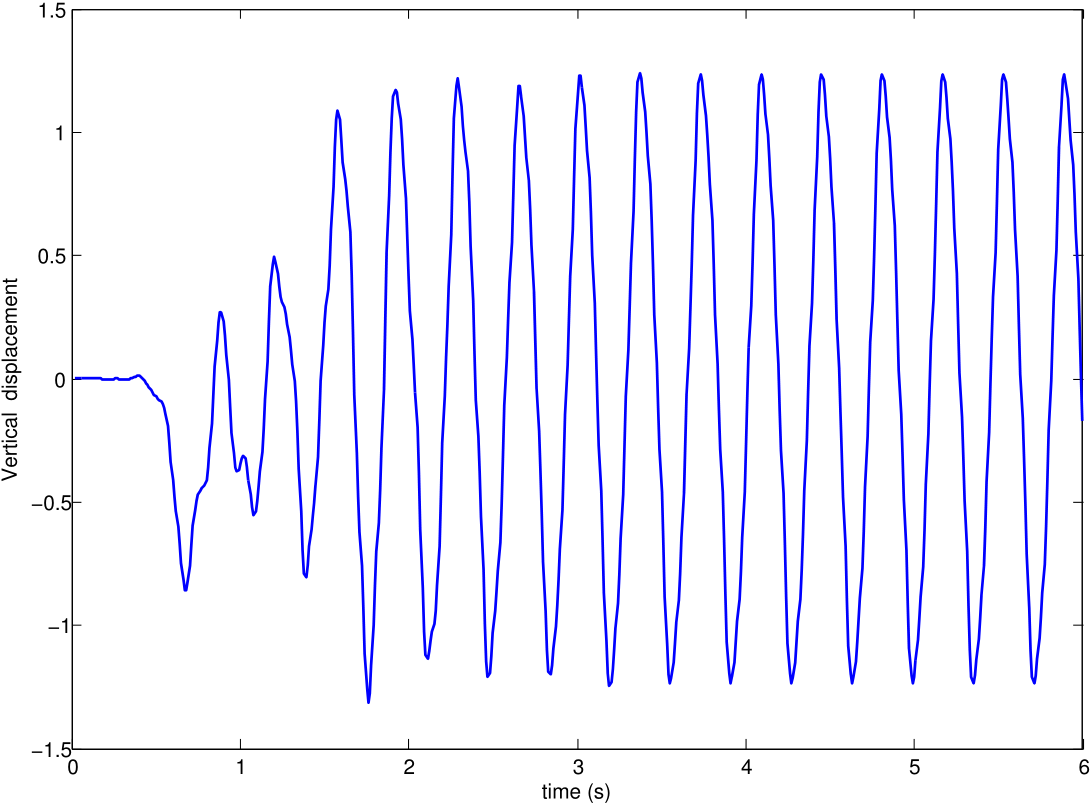}
		\caption*{\scriptsize (b) Explicit Taylor-Galerkin method.}		
	\end{minipage}     		
	\captionsetup{justification=centering}
	\caption {\scriptsize Displacement of leaflet tip as a function of time.} 
	\label{Displacement of leaflet tip as a function of time}	
\end{figure}

The evolution of the vertical displacement of the leaflet tip with respect to time is plotted in Figure \ref{Displacement of leaflet tip as a function of time}(a). Both the magnitude (1.34) and the frequency (2.94) have a good agreement with the result of \cite{wall1999fluid}, using a fitted ALE mesh and of \cite{Kadapa_2016}, using a monolithic unfitted mesh approach. The Taylor-Galerkin method is also tested which uses $\Delta t=2.0\times 10^{-4}s$ as a stable time step,  and a corresponding result is shown in \ref{Displacement of leaflet tip as a function of time}(b) which has a similar magnitude (1.24) and frequency (2.86). These results are all within the range of values in \cite[Table 4]{Kadapa_2016}. Note that since the initial condition before oscillation for these simulations is an unstable equilibrium, the first perturbation from this regime is due to numerical disturbances. Consequently, the initial transient regimes observed for the two methods (implicit Least-squares and explicit Taylor-Galerkin methods) are quite different. It is possible that an explicit method causes these numerical perturbations more easily, therefore makes the leaflet start to oscillate at an earlier stage than when using the implicit Least-squares approach.

\subsection{Solid disc in a cavity flow}
This numerical example is used to compare our UFEM with the IFEM, which is cited in \cite{Wang_2009,Zhao_2008}. In order to compare some details, we also implement the IFEM, but we implemented it on an adaptive mesh with hanging nodes, and we use the isoparametric FEM interpolation function rather than the discretized delta function or RKPM function of \cite{zhang2004immersed,Zhang_2007}. 

\begin{figure}[h!]
	\begin{minipage}[h!]{0.5\linewidth}
		\centering  
		\includegraphics[width=1.8in,angle=0]{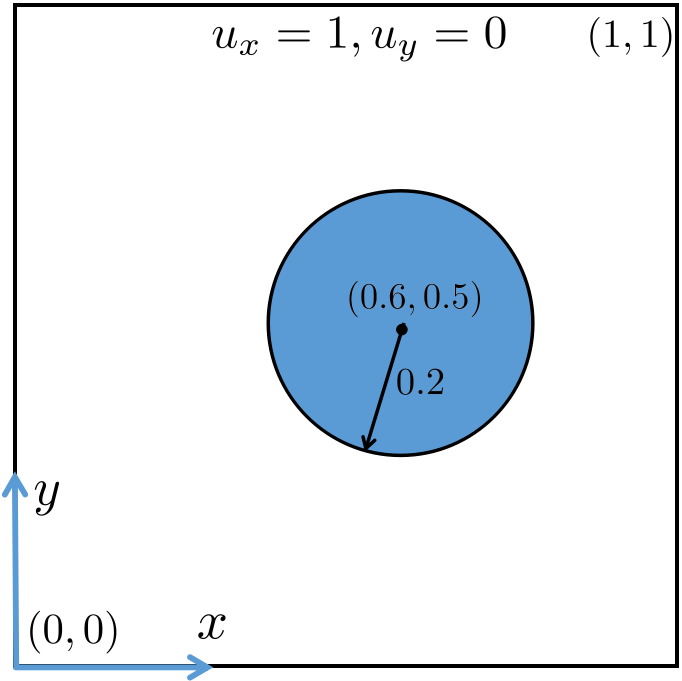}	
		\captionsetup{justification=centering}
		\caption {\scriptsize Computational domain for \\ cavity flow, taken from \cite{Zhao_2008}.} 
		\label{Computational domain for cavity flow}
	\end{minipage}
	\begin{minipage}[h!]{0.5\linewidth}
		\centering  
		\includegraphics[width=1.8in,angle=0]{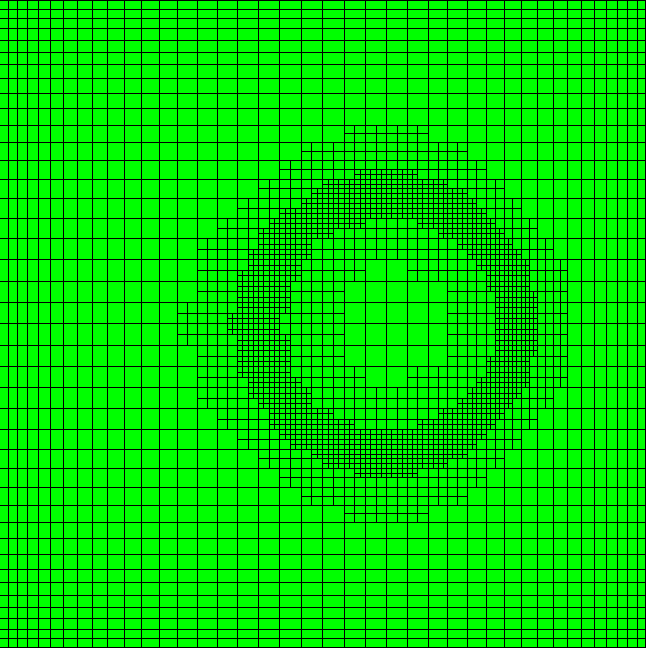}
		\captionsetup{justification=centering}	
		\caption {\scriptsize Adaptive mesh for \\ cavity flow.} 
		\label{Adaptive mesh for cavity flow}	
	\end{minipage}     			
\end{figure}

The fluid's density and viscosity are 1 and 0.01 respectively, and the following solid properties are chosen to undertake the tests: $\rho^s$=1 and $\mu^s$=0.1 or 1. The horizontal velocity on the top boundary of the cavity is prescribed as 1 and the vertical velocity is fixed to be 0 as shown in Figure \ref{Computational domain for cavity flow}. The velocities on the other three boundaries are all fixed to be 0, and pressure at the bottom-left point is fixed to be 0 as a reference point.

\begin{figure}[h!]
	\begin{minipage}[h!]{1\linewidth}
		\centering  
		\includegraphics[width=4in,angle=0]{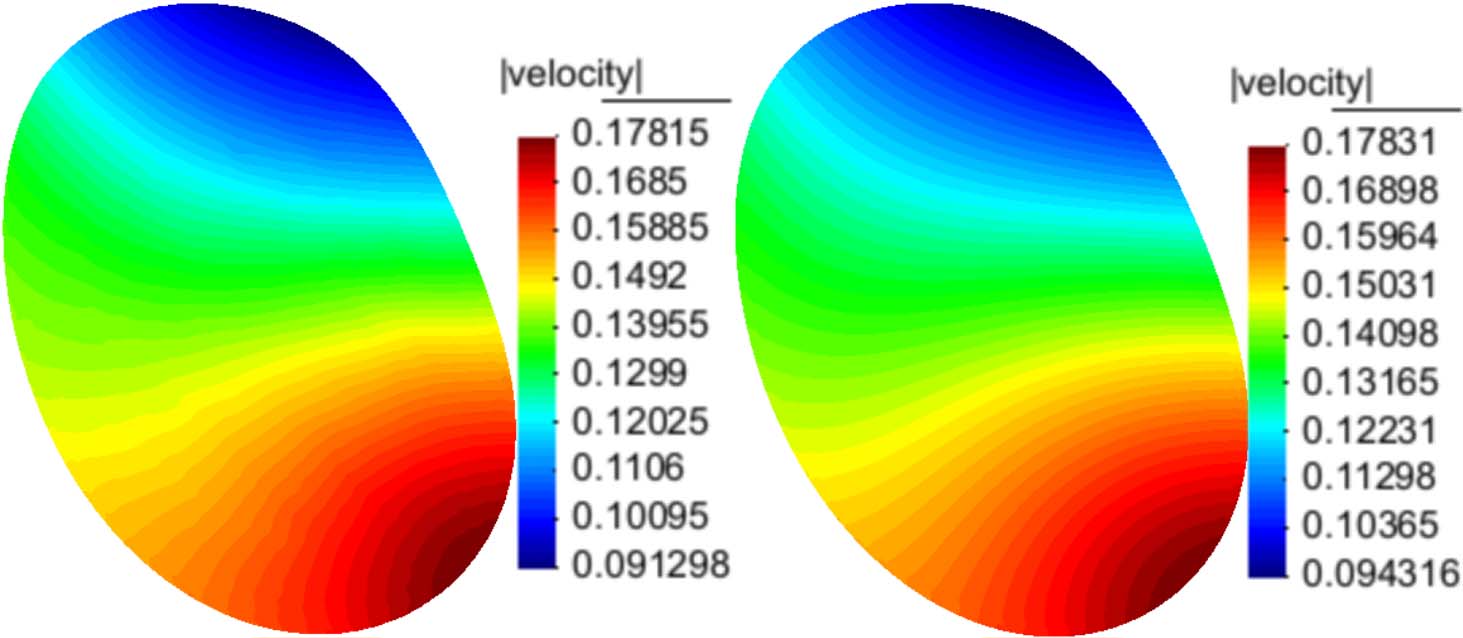}	
		\caption*{\scriptsize(a) $t=2.0s$.}
	\end{minipage}
	\begin{minipage}[h!]{1\linewidth}
		\centering  
		\includegraphics[width=4in,angle=0]{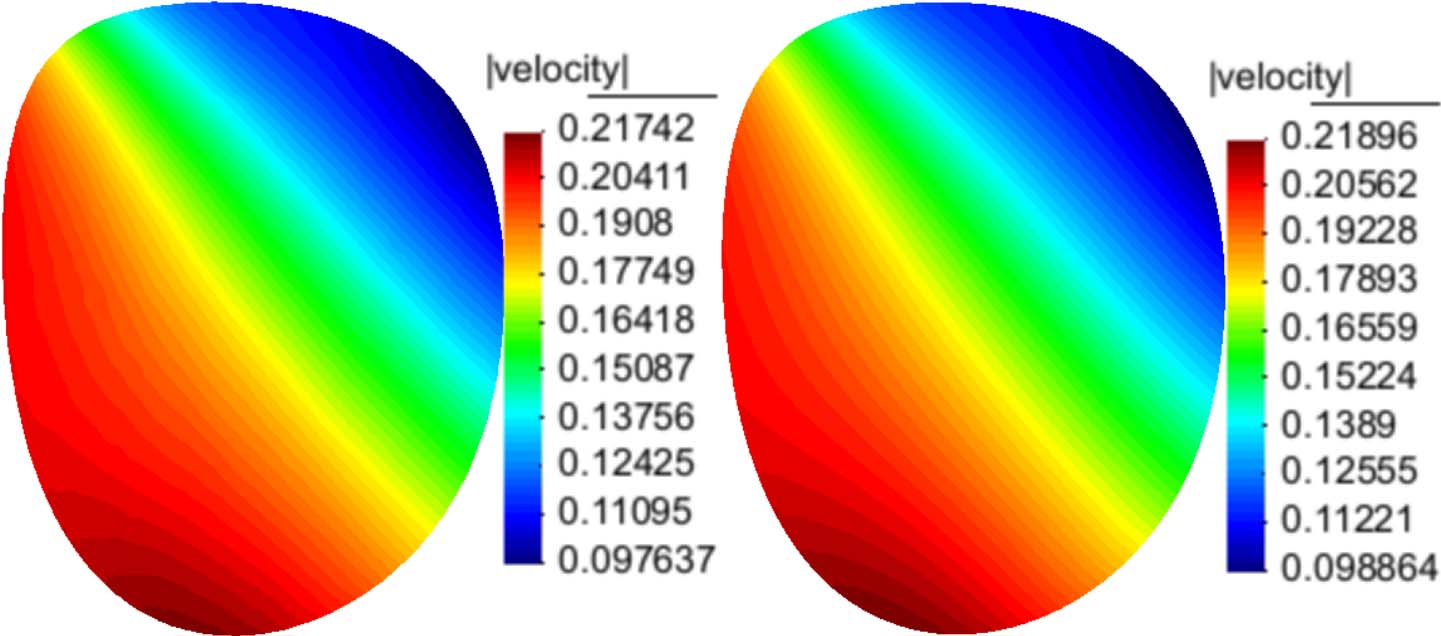}
		\caption*{\scriptsize(b) $t=3.0s$.}		
	\end{minipage}     		
	\captionsetup{justification=centering}
	\caption {\scriptsize Velocity norm for a soft solid $\left(\mu^s=0.1\right)$ in a driven cavity flow \\
		using UFEM (left) and IFEM (right).} 
	\label{ Velocity norm for a soft solid1}	
\end{figure}

\begin{figure}[h!]
	\begin{minipage}[h!]{1\linewidth}
		\centering  
		\includegraphics[width=4in,angle=0]{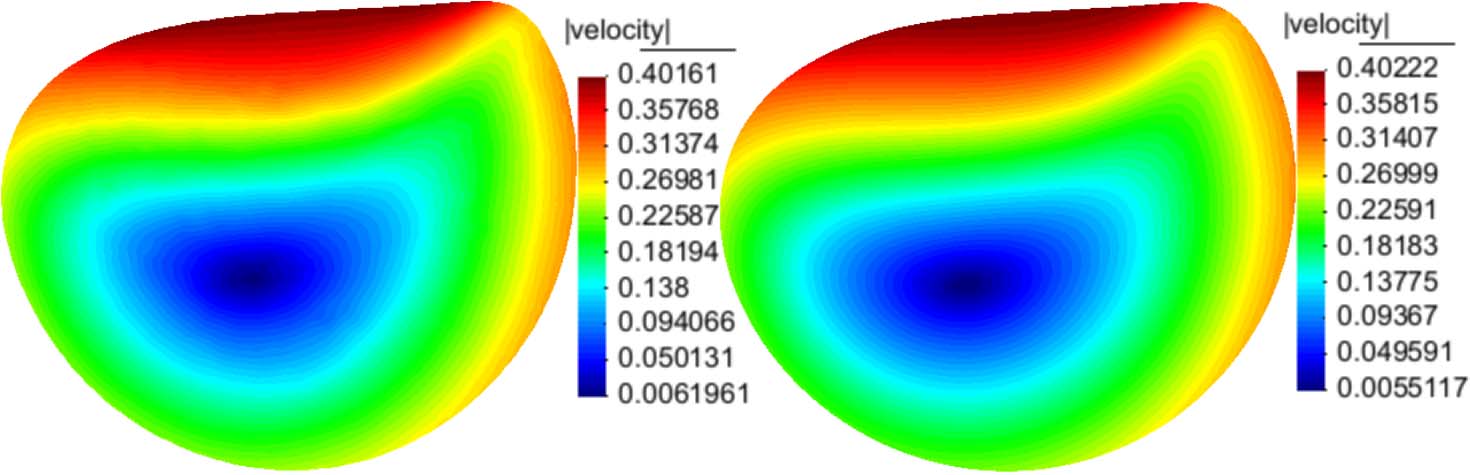}	
		\caption*{\scriptsize(a) $t=5.0s$.}
	\end{minipage}
	\begin{minipage}[h!]{1\linewidth}
		\centering  
		\includegraphics[width=4in,angle=0]{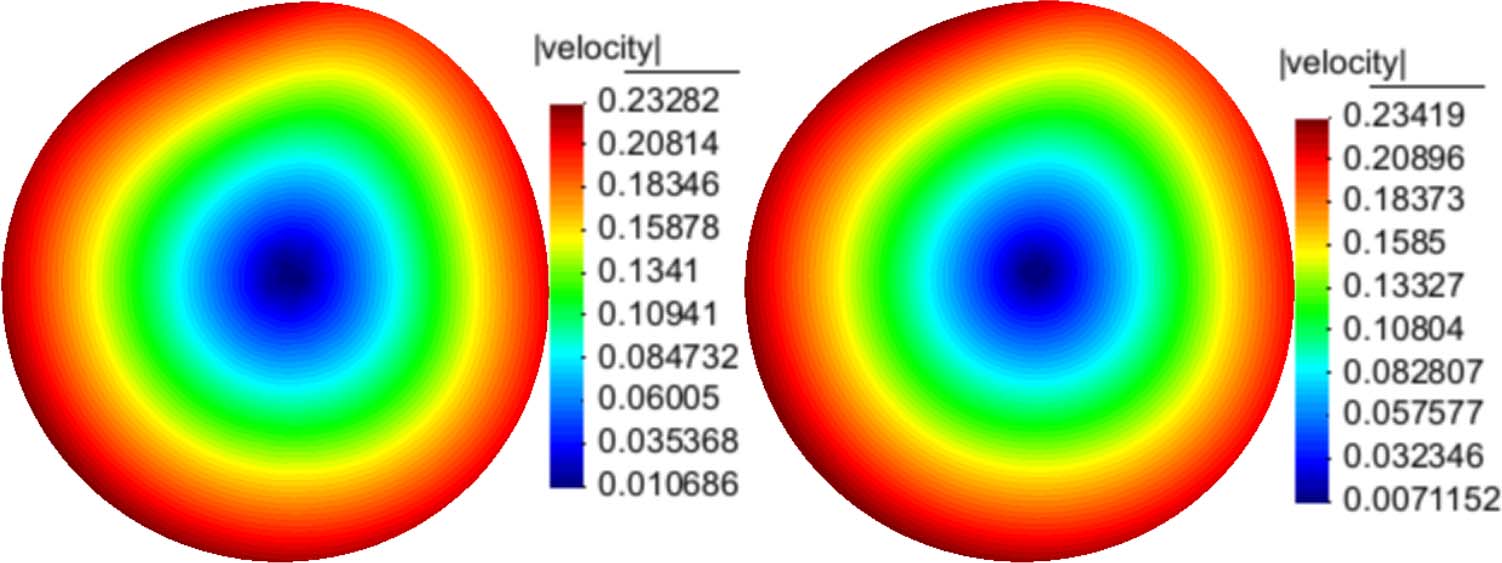}
		\caption*{\scriptsize(b) $t=25.0s$.}		
	\end{minipage}     		
	\captionsetup{justification=centering}
	\caption {\scriptsize Velocity norm for a soft solid $\left(\mu^s=1.0\right)$ in a driven cavity flow \\
		using UFEM (left) and IFEM (right), Least-squares method for convection step.} 
	\label{ Velocity norm for a soft solid2}	
\end{figure}

\begin{figure}[h!]
	\centering  
	\includegraphics[width=4in,angle=0]{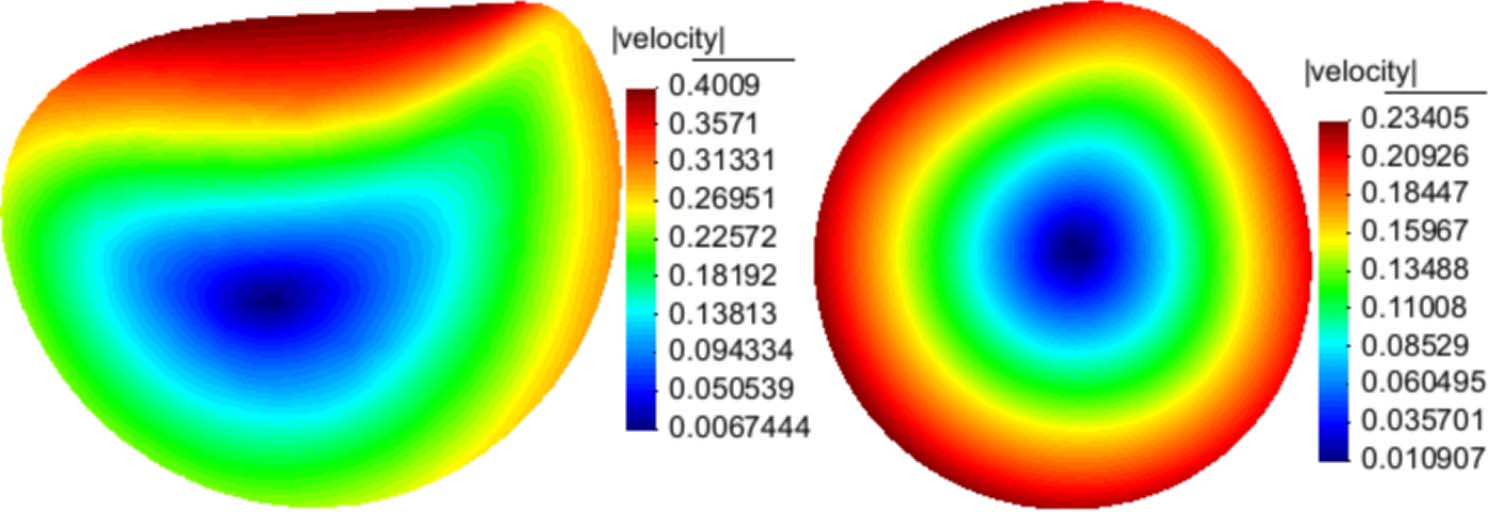}
	\caption* {\scriptsize (a) $t=5.0s \qquad\qquad\qquad\qquad\qquad\qquad$ (b) $t=25.0s$} 
	\caption {\scriptsize Velocity norm for $\mu^s=1.0$, Taylor-Galerkin method for convection step.} 
	\label{Velocity norm using TG}	
\end{figure}

In order to compare the UFEM and IFEM, we use the same meshes for fluid and solid: the solid mesh has 2381 nodes and the fluid mesh locally has a similar number of nodes (adaptive, see Figure \ref{Adaptive mesh for cavity flow}). First the implicit Least-squares method is used to solve the convection step, and the time step is $\Delta t=1.0\times10^{-3}$. Figure \ref{ Velocity norm for a soft solid1} and Figure \ref{ Velocity norm for a soft solid2} show the configuration of the disc at different stages, from which we do not observe significant differences of the velocity norm even for a long run as shown in Figure \ref{ Velocity norm for a soft solid2} (b). Then the explicit Taylor-Galerkin method is tested, and we achieve almost the same accuracy by using the same time step. The magnitudes of velocity at the same stages of Figure \ref{ Velocity norm for a soft solid2} are presented in Figure \ref{Velocity norm using TG}.

We should mention that for the case $\mu^s=0.1$, as the disc arrives at the top of the cavity (time $>3.0$) the quality of the solid mesh does begin to deteriorate using our UFEM. We do not currently seek to improve the mesh quality (using an arbitrary Lagrangian-Eulerian (ALE) update \cite{peterson1999solution}, for example) however this would be necessary in order to reduce the shear modulus further without compromising the quality of the solid mesh.

Conversely, a large $\mu^s$ makes the solid behave like a rigid body. For the proposed UFEM, we can use $\mu^s=100$ or larger without changing the time step, whereas for the IFEM the simulation always breaks down for $\mu^s=100$, however small the time step, due to the huge FSI force on the right-hand side of the FSI system.

\subsection{Solids in a channel with gravity}
We first simulate a falling disc due to gravity in order to further validate the accuracy of the UFEM. We then show a simulation of the evolution of different shapes of solids falling and rising in a channel in order to show the flexibility and robustness of the proposed UFEM. 

\begin{figure}[h!]
	\begin{minipage}[h!]{0.5\linewidth}
	\centering  
	\includegraphics[width=1.2in,angle=0]{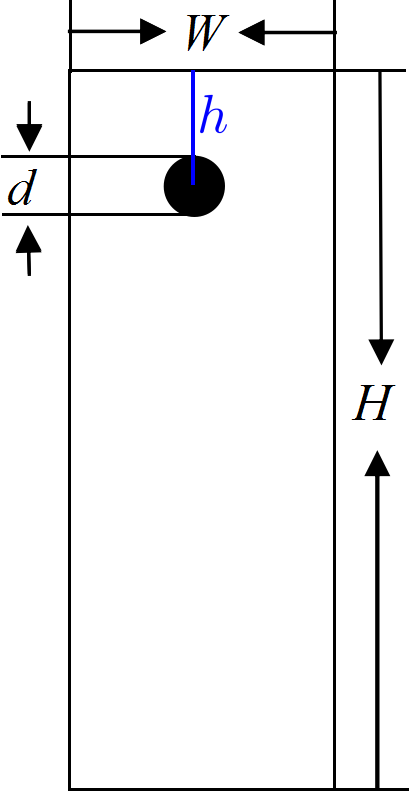}
	\captionsetup{justification=centering}
	\caption {\scriptsize Computational domain for \\ a falling disc.} 
	\label{Computational domain for a falling disc}	
	\end{minipage}
	\begin{minipage}[h!]{0.5\linewidth}
	\centering  
	\includegraphics[width=1in,angle=0]{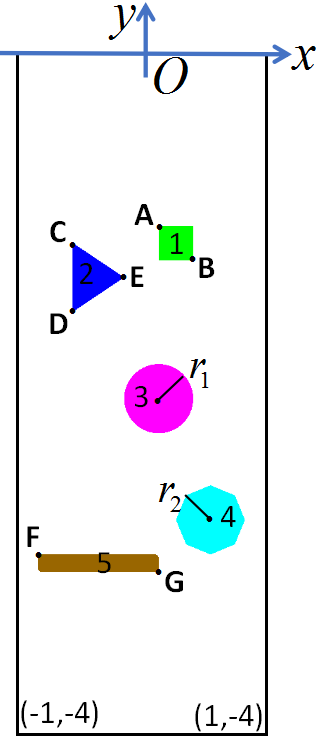}
	\captionsetup{justification=centering}
	\caption {\scriptsize Computational domain for different shapes of solids with different properties.} 
	\label{Computational domain for different shapes of solids with different properties}		
	\end{minipage}     		
\end{figure}

The test of a falling disc in a channel is cited by \cite{Zhang_2007,Hesch_2014} in order to validate the IFEM and a monolithic method respectively. The computational domain and parameters are illustrated in Figure \ref{Computational domain for a falling disc} and Table \ref{Fluid and material properties of a falling disc} respectively. The fluid velocity is fixed to be 0 on all boundaries except the top one.

\begin{table}[h!]
	\centering
	\begin{tabular}{|c|c|}
		\hline
		Fluid & Disc \\
		\hline
		$W=2.0$ $cm$ & $d=0.0125$ $cm$ \\
		$H=4.0$ $cm$ & $h=0.5$ $cm$ \\
		$\rho^f=1.0$ $\left.g\right/{cm^3}$ & $\rho^s=1.2$ $\left.g\right/{cm^3}$ \\
		$\mu^f=1.0$ $\left.dyne\cdot s\right/{cm^2}$ &  $\mu^s=10^8$ $\left.dyne\right/{cm^2}$ \\
		$g=980$ $\left.cm\right/s^2$ & $g=980$ $\left.cm\right/s^2$ \\		
		\hline
	\end{tabular}
	\caption{Fluid and material properties of a falling disc.}
	\label{Fluid and material properties of a falling disc}
\end{table}

There is also an empirical solution of a rigid ball falling in a viscous fluid \cite{Hesch_2014}, for which the terminal velocity, $u_t$, under gravity is given by
\begin{equation}\label{empirical formula}
u_t=\frac{\left(\rho^s-\rho^f\right)gr^2}{4\mu^f}\left(ln\left(\frac{L}{r}\right)-0.9157+1.7244\left(\frac{r}{L}\right)^2-1.7302\left(\frac{r}{L}\right)^4\right),
\end{equation}
where $\rho^s$ and $\rho^f$ are the density of solid and fluid respectively, $\mu^f$ is viscosity of the fluid, $g=980$ $\left.cm\right/s^2$ is acceleration due to gravity, $\left.L=W\right/2$ and $r$ is the radius of the falling ball. We choose $\mu^s=10^8$ $\left.dyne\right/cm^2$ to simulate a rigid body here, and $\mu^s=10^{12}$ $\left.dyne\right/cm^2$ is also applied, which gives virtually identical result.

\begin{figure}[h!]
	\centering  
	\includegraphics[width=4in,angle=0]{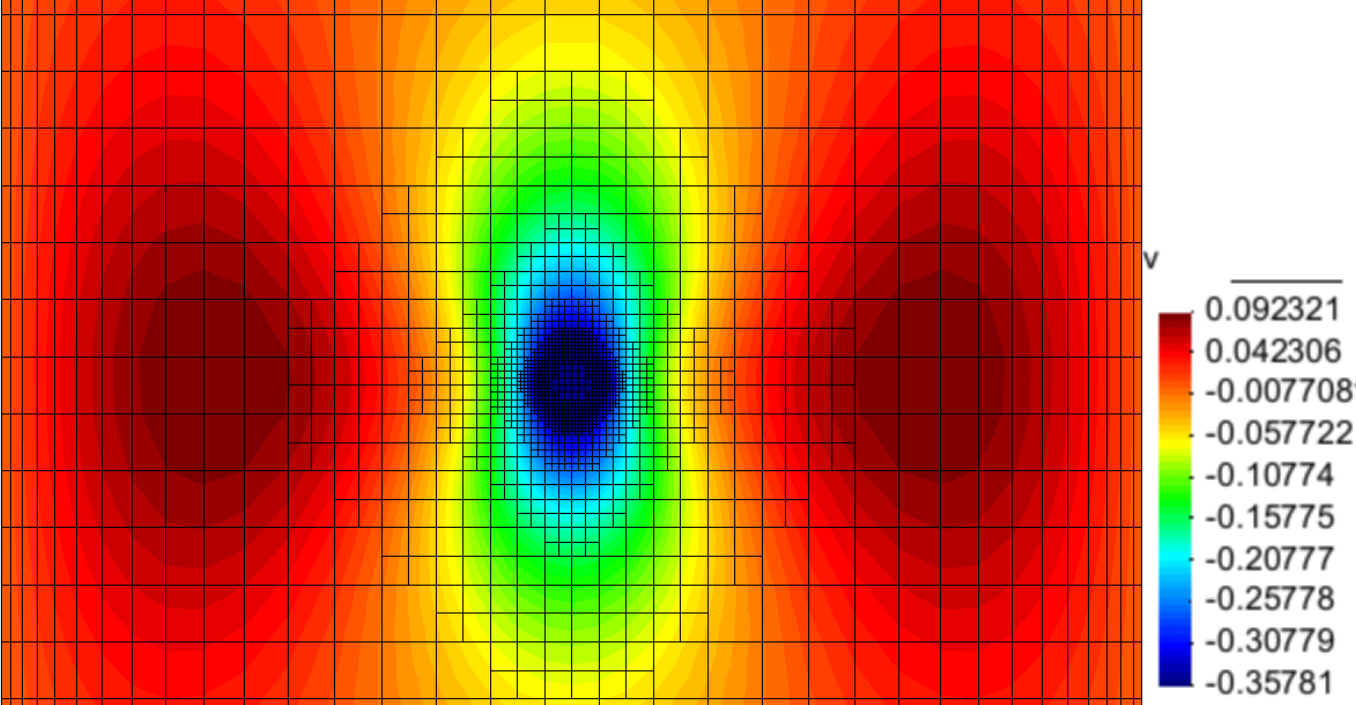}
	\captionsetup{justification=centering}
	\caption {\scriptsize Contour of vertical velocity at $t=1s$ (fine mesh).} 
	\label{Contour of vertical velocity at t=1s (fine mesh)}						
\end{figure}

\begin{figure}[h!]
	\centering  
	\includegraphics[width=4in,angle=0]{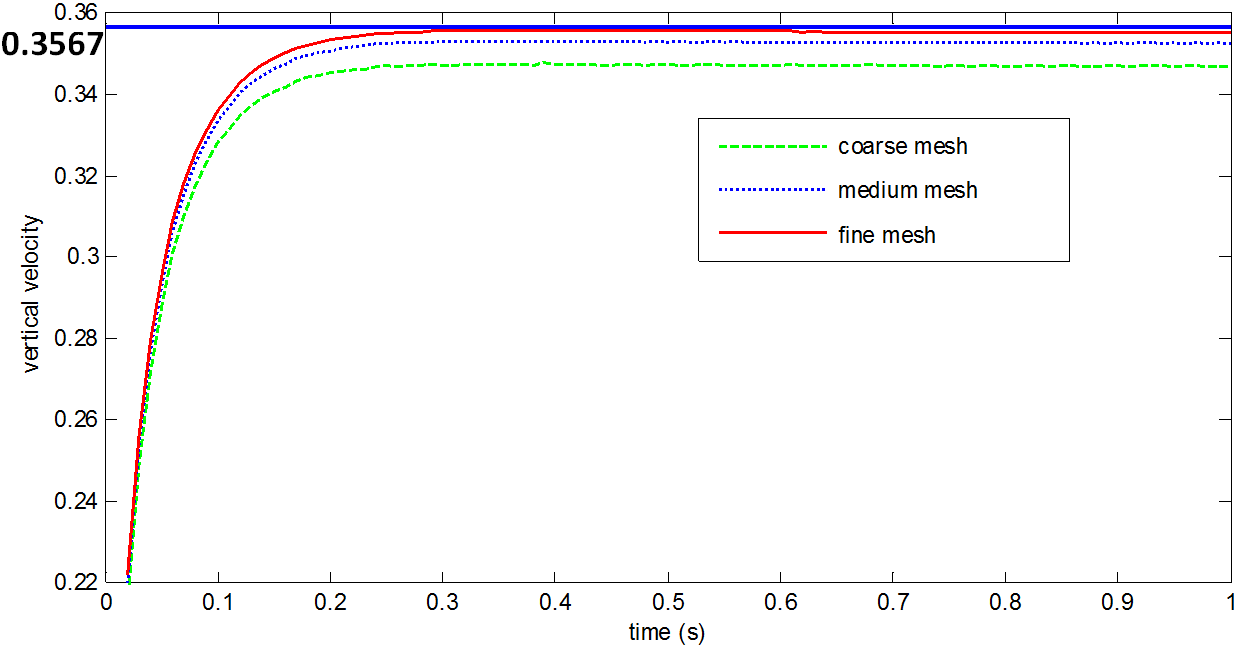}
	\captionsetup{justification=centering}
	\caption {\scriptsize Evolution of velocity at the center of a falling disc. \\
		(The blue solid line represents the empirical solution from formula (\ref{empirical formula}).)} 
	\label{Evolution of velocity at center a falling disc}						
\end{figure}

Three different meshes are used: the disc boundary is represented with 28 nodes (coarse), 48 nodes (medium), or 80 nodes (fine). The fluid mesh near the solid boundary has the same mesh size, and a stable time step $t=0.005s$ is used for all the three cases. The Least-squares method is used to treat the convection step in all these tests. A local snapshot of the vertical velocity with the adaptive mesh is shown in Figure \ref{Contour of vertical velocity at t=1s (fine mesh)}. From the fluid velocity pattern around the disc, we can observe that the disc behaves like a rigid body as expected. In addition, the evolution of the velocity of the mid-point of the disc is shown in Figure \ref{Evolution of velocity at center a falling disc}, from which it can be seen that the numerical solution converges from below to the empirical solution.

Reference  \cite{Hesch_2014} uses a monolithic method to simulate multiple rigid and deformable discs in a gravity channel. We have implemented this example and obtain very similar results. Rather than replicate these here however, we instead show a more complex example, as illustrated in Figure \ref{Computational domain for different shapes of solids with different properties}. The computational domain, boundary conditions and the fluid properties are the same as the above one-disc test. All the solids are numbered at their initial positions as shown in Figure \ref{Computational domain for different shapes of solids with different properties} with $A(0,-1)$, $B(0.2,-1.2)$, $C(-0.5,-1.1)$, $D(-0.5,-1.5)$, $E(-0.2,-1.3)$, $F(-0.7,-2.9)$ and $G(0,-3)$. The center and radius $\left(r_1\right)$ of the $3^{rd}$ solid (circle) are $(0,-2)$ and $0.2$ respectively, and the center and radius $\left(r_2\right)$ the $4^{th}$ solid (octagon) are $(0.3,-2.7)$ and $0.2$ respectively. The solid properties are illustrated in Table \ref{Solids properties in Figure}.

\begin{table}[h!]
	\centering
	\begin{tabular}{|c|c|c|}
		\hline
		No. of solid & Density $\left(\left.g\right/cm^3\right)$ & Shear modulus $\left(\left.dyne\right/{cm^2}\right)$ \\
		\hline
		1 & 1.3 & $10^4$ \\
		\hline
		2 & 1.2 & $10^3$ \\		
		\hline
		3 & 1.0 & $10$ \\
		\hline
		4 & 0.8 & $10^6$ \\				
		\hline
		5 & 0.7 & $10^2$ \\								
		\hline
	\end{tabular}
	\caption{Properties for multi-solids falling in a channel as shown in Figure \ref{Computational domain for different shapes of solids with different properties}.}
	\label{Solids properties in Figure}
\end{table}

\begin{figure}[h!]
	\begin{minipage}[h!]{0.5\linewidth}
		\centering  
		\includegraphics[width=2in,angle=0]{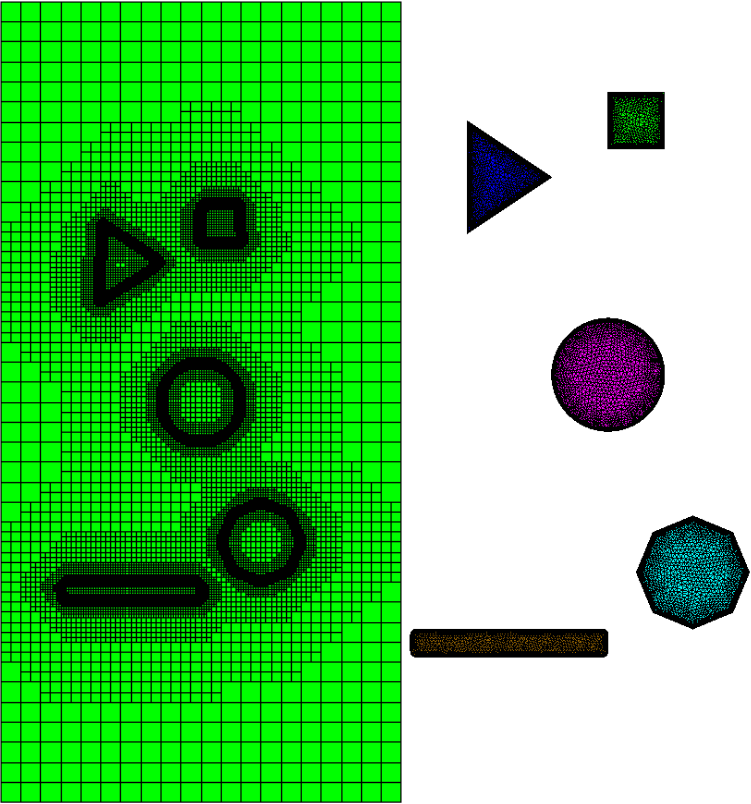}
		\captionsetup{justification=centering}
		\caption* {\scriptsize (a) $t=0.0$} 
	\end{minipage}
	\begin{minipage}[h!]{0.5\linewidth}
		\centering  
		\includegraphics[width=1.6in,angle=0]{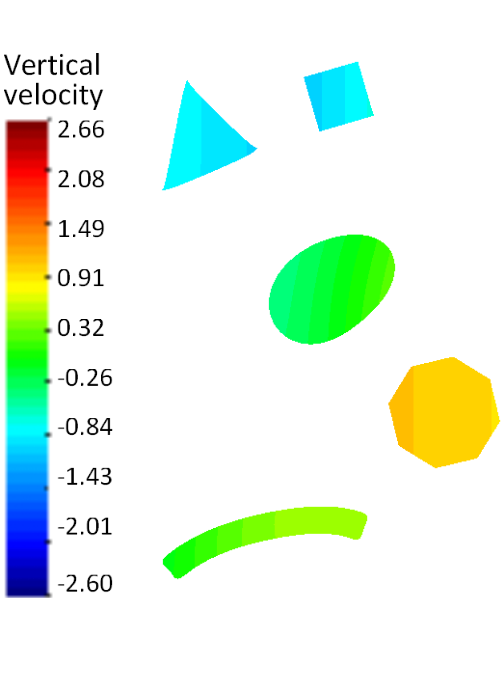}
		\captionsetup{justification=centering}
		\caption* {\scriptsize (b) $t=0.3$} 
	\end{minipage}
	\begin{minipage}[h!]{1\linewidth}
		\centering  
		\includegraphics[width=4in,angle=0]{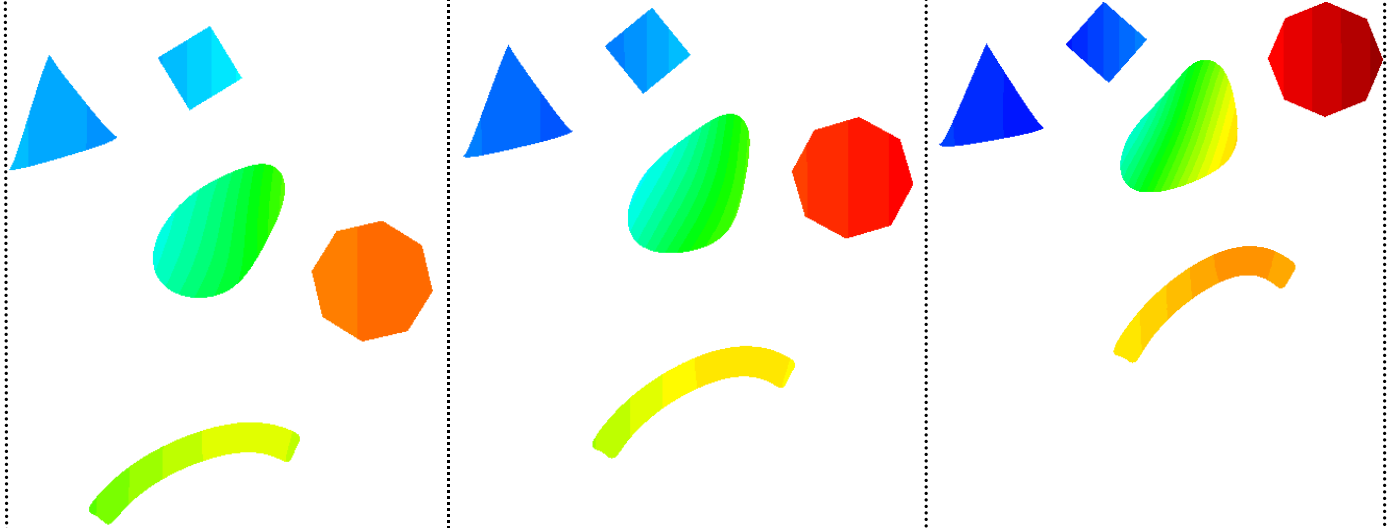}
		\captionsetup{justification=centering}
		\caption* {\scriptsize (c) $t=0.5 \qquad\qquad\qquad\qquad$ (d) $t=0.6 \qquad\qquad\qquad\qquad$ (e) $t=0.7$} 
	\end{minipage}
	\begin{minipage}[h!]{1\linewidth}
		\centering  
		\includegraphics[width=4in,angle=0]{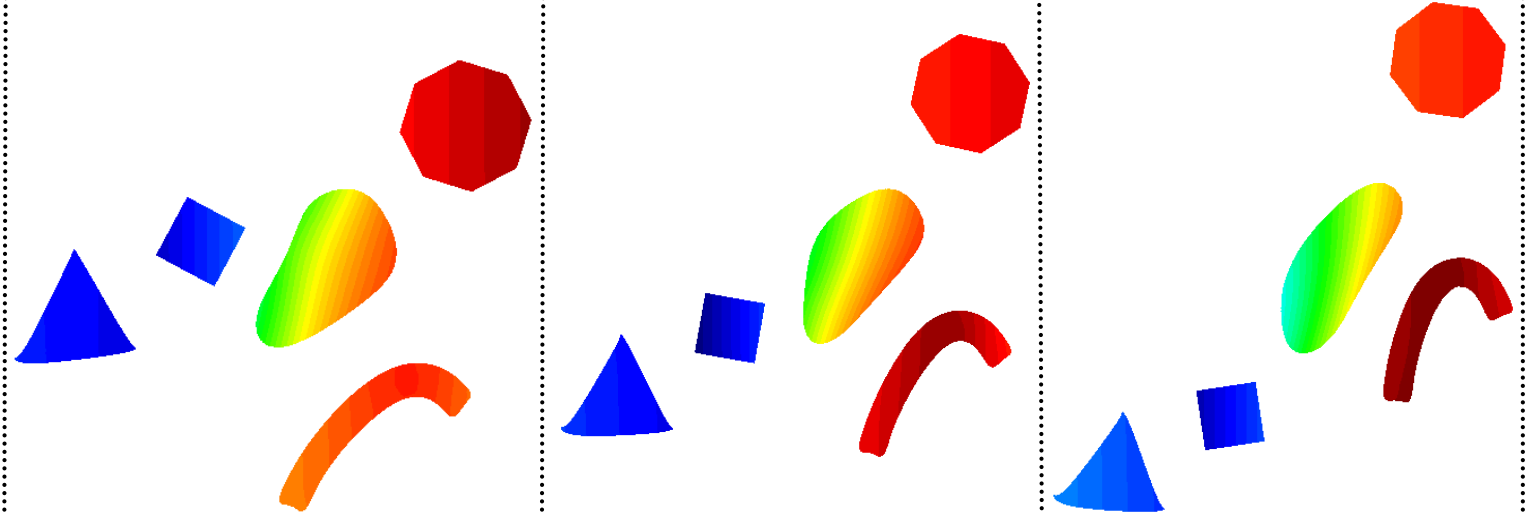}
		\captionsetup{justification=centering}
		\caption* {\scriptsize (f) $t=0.8 \qquad\qquad\qquad\qquad$ (g) $t=0.9 \qquad\qquad\qquad\qquad$ (h) $t=1.0$} 
	\end{minipage}   	
	\captionsetup{justification=centering}
	\caption {\scriptsize Contours of vertical velocity at different times.} 
	\label{Contours of vertical velocity at different times}				
\end{figure}

\begin{figure}[h!] 	
	\begin{minipage}[h!]{1\linewidth}
		\centering  
		\includegraphics[width=4in,angle=0]{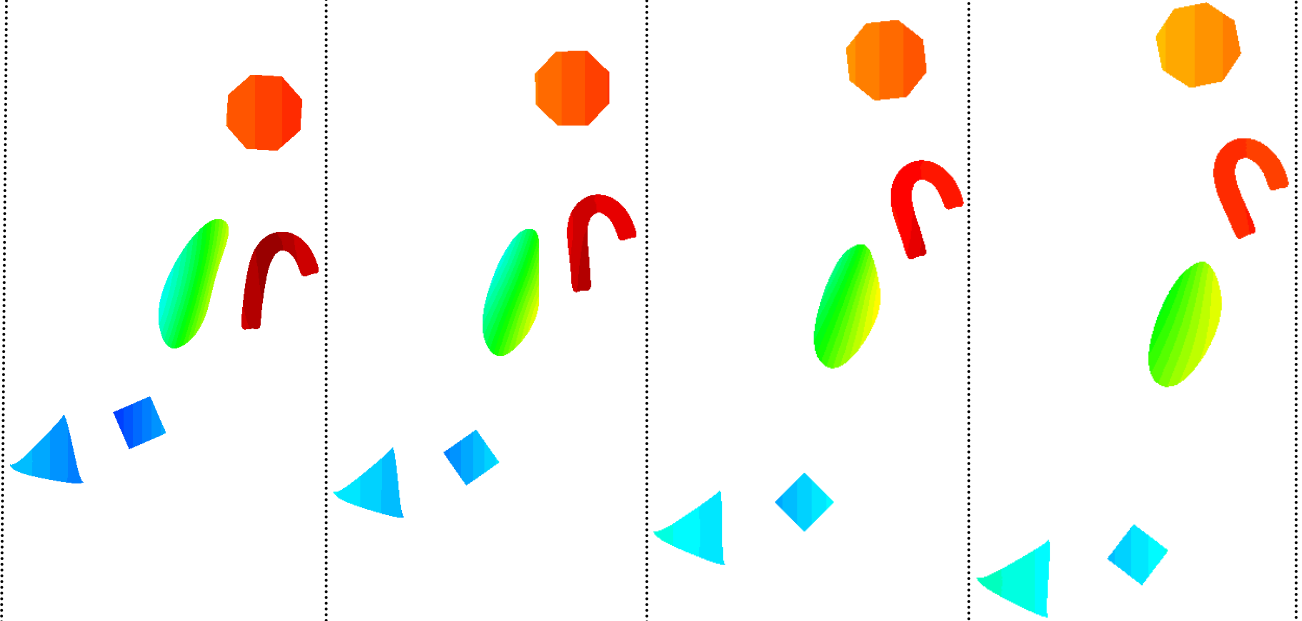}
		\captionsetup{justification=centering}
		\caption* {\scriptsize (i) $t=1.1 \qquad\qquad $ (j) $t=1.2 \qquad\qquad$ (k) $t=1.3 \qquad\qquad$ (l) $t=1.4$} 
	\end{minipage}      	
	\begin{minipage}[h!]{1\linewidth}
		\centering  
		\includegraphics[width=4in,angle=0]{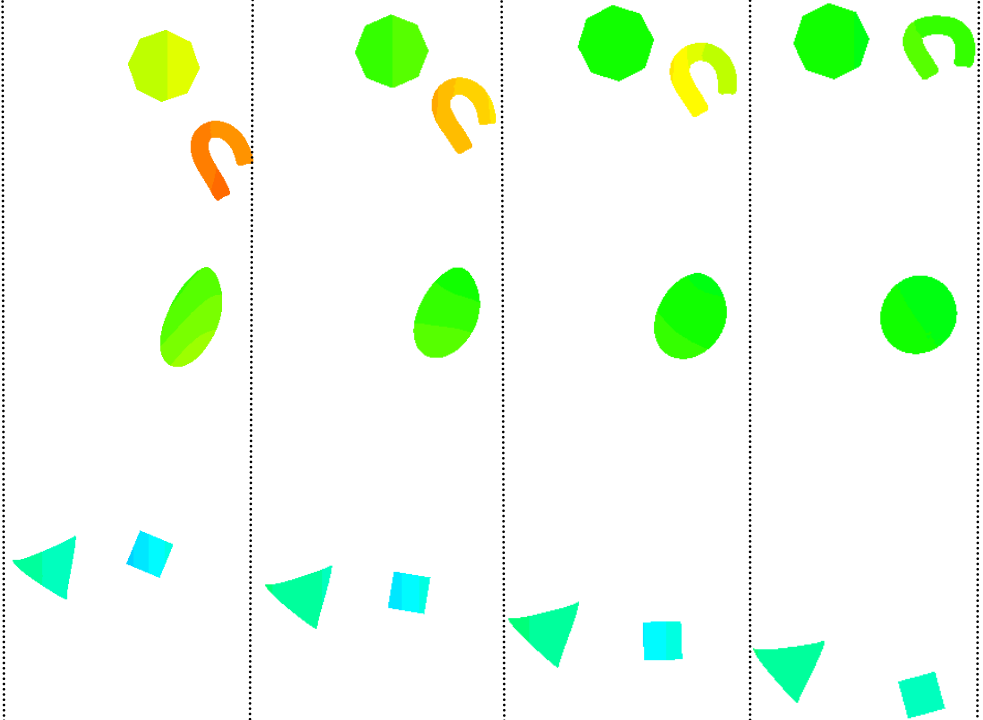}
		\captionsetup{justification=centering}
		\caption* {\scriptsize (m) $t=1.6 \qquad\qquad $ (n) $t=1.8 \qquad\qquad $ (o) $t=2.0 \qquad\quad$ (p) $t=2.4$} 
	\end{minipage}  			
	\captionsetup{justification=centering}
	\caption {{\scriptsize Contours of vertical velocity at different times} (continued).} 	
	\label{Contours of vertical velocity at different times(continued)}	
\end{figure}

A high resolution of each solid boundary is used in this simulation as shown in Figure \ref{Contours of vertical velocity at different times} (a),  which can guarantee the mesh quality during the whole process of evolution, and a stable time step $t=0.002s$ is used. Snapshots of the solids at different times are shown in Figure \ref{Contours of vertical velocity at different times} and \ref{Contours of vertical velocity at different times(continued)}.

\section{Conclusion and future works}
In this article we introduce a new unified finite element method (UFEM) for fluid-structure interaction, which can be applied to a wide range of problems, from small deformation to very large deformation and from very soft solids through to very rigid solids. Several numerical examples, which are widely used in the literature of Immersed FEM and monolithic methods, are implemented to validate the proposed UFEM. 

The UFEM combines features from the IFEM and from monolithic methods. Nevertheless, it differs from each of them in the following aspects. Firstly, UFEM is a semi-explicit (explicitly linearizing the constitutive equation of solid and implicitly coupling FSI interaction) scheme, similar to IFEM, however UFEM solves the solid equations and fluid equations together while the classical IFEM does not solve the solid equations; secondly, both UFEM and monolithic methods solve solid equations, however UFEM solves one velocity field in the solid domain using FEM interpolation, while monolithic methods solve one velocity field and one displacement field in the solid domain using Lagrangian multipliers. In summary therefore we believe that UFEM has the potential to offer the robustness and range of operation of monolithic methods, but at a computational cost that is much closer to that of the immersed finite element methods.

The following generalizations of our proposed UFEM approach will be considered in the future: (1) Implementation in 3D using adaptive mesh with hanging nodes; (2) implementation for non-Newtonian flow; (3) an efficient preconditioned iterative solver for the UFEM algebraic system; (4) a second order splitting scheme in time.

\appendix
\section{A method to treat hanging nodes}
An adaptive mesh with hanging nodes reduces the number of degrees of freedom compared to uniform refinement, hence, decreases the cost of computation. However, the nature of hanging nodes has the potential to cause discontinuity and breaks the framework of the finite element shape functions, which, therefore, needs special treatment in  finite element codes.

In order to treat the hanging nodes, one can construct a conforming shape function \cite{Gupta_1978,Fries_2010} or constrain and cancel the degree of freedom at the hanging nodes \cite{Fries_2010,Bangerth_2009}. The former is very appealing and leads to optimal convergence, but it is difficult to extend to high-order shape functions \cite{Zander_2015}. In this article we will adopt the latter method and only use 2-level hanging nodes, which means at most 2 hanging nodes are allowed in one element (this can be guaranteed by imposing safety layers to ensure that neighbouring element nodes differ by more than one level of refinement). The implementation of arbitrary-level hanging nodes can be found in \cite{Zander_2015,_ol_n_2008,Ooi_2015}.

\begin{figure}[h!] 	
	\begin{minipage}[h!]{0.5\linewidth}
		\centering  
		\includegraphics[width=2in,angle=0]{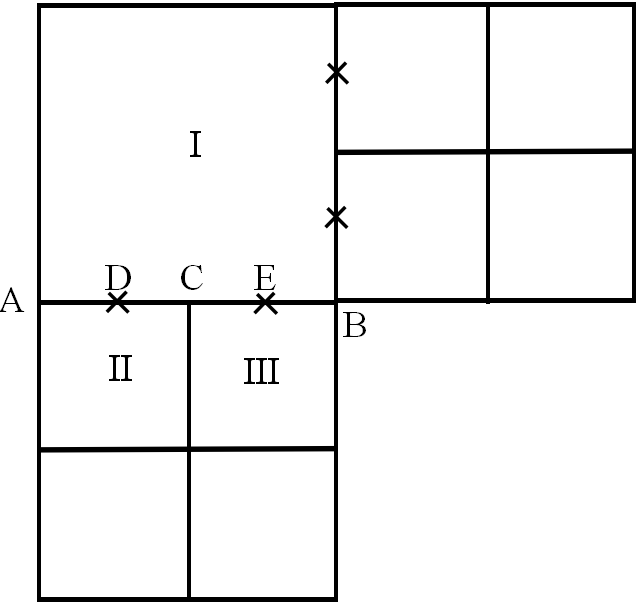}
		\captionsetup{justification=centering}
		\caption* {Figure A.1: Elements with hanging nodes.} 
	\end{minipage}      	
	\begin{minipage}[h!]{0.5\linewidth}
		\centering  
		\includegraphics[width=2in,angle=0]{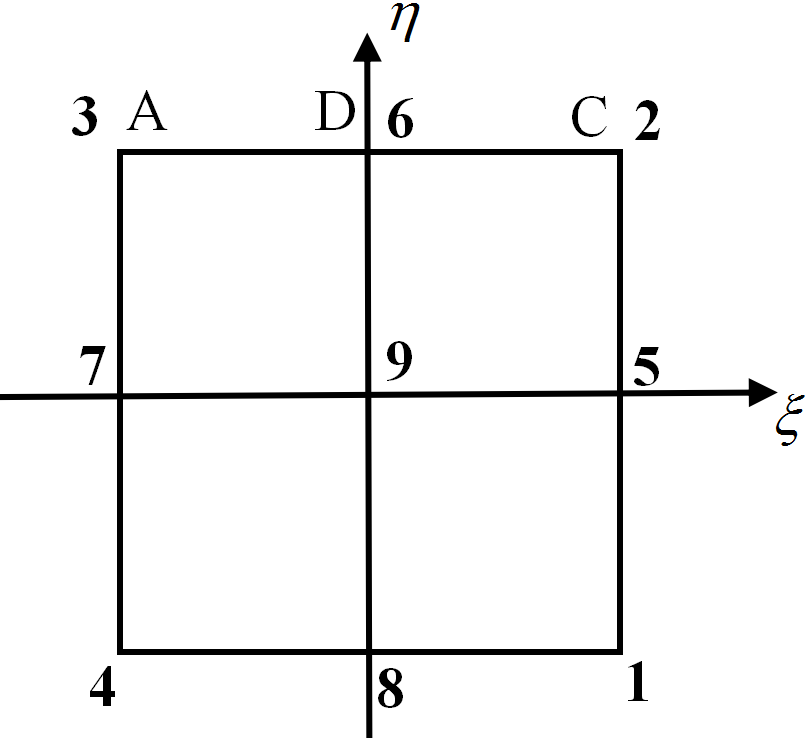}
		\captionsetup{justification=centering}
		\caption* {Figure A.2: Element II in Figure A.1 \\ in the reference coordinate system.} 
	\end{minipage}  					
\end{figure}

For a quadrilateral element, when the velocity is interpolated by biquadratic shape functions and the pressure is interpolated by bilinear shape functions, the implementation of hanging nodes must be different for each, as shown in Figure A.1.

For example, when velocity is interpolated, point $D$ is a hanging node for element II, and point $E$ is a hanging node for element III. When pressure is interpolated, point $C$ is a hanging node for both the element II and III. Take element II for example, if we use the constraint method to cancel the hanging node’s degree of freedom, then

\begin{equation}\label{eqa1}
u_i^D=\frac{3}{8}u_i^A-\frac{1}{8}u_i^B+\frac{3}{4}u_i^C \quad i=\left(1,2\right)
\end{equation}

\begin{equation}\label{eqa2}
p^C=\frac{1}{2}p^A+\frac{1}{2}p^B
\end{equation}
where $u_i$ and $p$ are velocity components and pressure respectively defined at the corresponding nodes. The interpolation coefficients can be calculated by putting edge $AB$ in a one dimensional finite element reference coordinate system.

Notice that when computing the element matrix II, point $B$ is outside of the element, but the element matrix II still contributes to node $B$ because of the hanging node $D$. So we can treat the two points, $B$ and $D$, as a master-slave couple, which means letting them share the same equation number in the final global linear equation system. However one should modify the element matrix II according to (\ref{eqa1}) and (\ref{eqa2}) in the following way before assembling it to the global matrix.

Suppose the element II is enumerated in the reference coordinate system as shown in Figure A.2. Then, formulae (\ref{eqa1}) and (\ref{eqa2}) imply the following equations:

\begin{equation}\label{interpolation for v}
\left(\begin{array}{c} u_i^1 \\ u_i^2 \\  u_i^3 \\ u_i^4 \\ u_i^5 \\ u_i^6 \\ u_i^7 \\ u_i^8 \\ u_i^9 \end{array} \right)
={\bf D}^v\left(\begin{array}{c} u_i^1 \\ u_i^2 \\  u_i^3 \\ u_i^4 \\ u_i^5 \\ u_i^B \\ u_i^7 \\ u_i^8 \\ u_i^9 \end{array} \right),
\quad
{\bf D}^v=
\begin{bmatrix}
{1} &  {}  & {}   & {}   & {}   & {}   & {}   & {}   & {}\\
{}  &  {1} & {}   & {}   & {}   & {}   & {}   & {}   & {}\\
{}  &  {}  & {1}  & {}   & {}   & {}   & {}   & {}   & {}\\
{}  &  {}  & {}   & {1}  & {}   & {}   & {}   & {}   & {}\\
{}  &  {}  & {}   & {}   & {1}  & {}   & {}   & {}   & {}\\
{}  &  {\frac{3}{4}}  & {\frac{3}{8}}   & {}   & {}   & {-\frac{1}{8}}   & {}   & {}   & {}\\
{}  &  {}  & {}   & {}  & {}   & {}   & {1}   & {}   & {}\\
{}  &  {}  & {}   & {}   & {}  & {}   & {}   & {1}   & {}\\
{}  &  {}  & {}   & {}   & {}  & {}   & {}   & {}   & {1}\\
\end{bmatrix}
.
\end{equation}

\begin{equation}\label{interpolation for p}
\left(\begin{array}{c} p^1 \\ p^2 \\  p^3 \\ p^4 \end{array} \right)
={\bf D}^p\left(\begin{array}{c} p^1 \\ p^B \\  p^3 \\ p^4 \end{array} \right),
\quad
{\bf D}^p=
\begin{bmatrix}
{1} &  {}  & {}   & {}\\
{}  &  {\frac{1}{2}} & {\frac{1}{2}}   & {} \\
{}  &  {}  & {1}  & {} \\
{}  &  {}  & {}   & {1} \\
\end{bmatrix}
.
\end{equation}

One should use matrices ${\bf D}^v$ and ${\bf D}^p$ to modify the element matrix II. Suppose ${\bf K}_e$
is the stiffness matrix of element II without consideration of hanging nodes, and the unknowns are arranged in the following column vector.
\begin{equation}\label{arrange dofs}
\left(u_1^1, u_1^2, \cdots u_1^9,v_1^1,v_1^2 \cdots v_1^9,p^1,p^2\cdots p^4 \right)^{\rm T}.
\end{equation}
It is clear that ${\bf K}_e=\left[k_{ij}\right]$ is a n$\times$n (n=22) matrix, and it could be modified by the following pseudocode, which distribute the contribution of hanging nodes to the corresponding nodes according to formula (\ref{eqa1}).

\begin{table}[h!]
\centering
\begin{tabular}{|l|l|l|}
	\hline
    for j=1 to n  &  for j=1 to n &  for j=1 to n  \\
    $k_{i_1j}=k_{i_1j}+k_{i_0j}\cdot 3/8$ &  $k_{ji_1}=k_{ji_1}+k_{ji_0}\cdot 3/8$  & $k_{i_0j}=-k_{i_0j}/8$ \\
    $k_{i_2j}=k_{i_2j}+k_{i_0j}\cdot 3/4$ &  $k_{ji_2}=k_{ji_2}+k_{ji_0}\cdot 3/4$  & $k_{ji_0}=-k_{ji_0}/8$ \\
    end  & end & end \\
	\hline
\end{tabular}
\end{table}

Let $i_0=6$, $i_1=3$, and $i_2=2$ (based on (\ref{arrange dofs})), sequentially executing the above three pieces of codes would modify the matrix ${\bf K}_e$ corresponding to the first component of velocity, and let $i_0=15$, $i_1=12$, and $i_2=11$ (based on (\ref{arrange dofs})), executing the above codes would modify the matrix ${\bf K}_e$ corresponding to the second component of velocity. Similarly, in order to modify the matrix corresponding to pressure, one can execute the following codes which are based on formula (\ref{eqa2}):

\begin{table}[h!]
	\centering
	\begin{tabular}{|l|l|l|}
		\hline
		for j=1 to n  &  for j=1 to n &  for j=1 to n  \\
		$k_{i_1j}=k_{i_1j}+k_{i_2j}/2$ &  $k_{ji_1}=k_{ji_1}+k_{ji_2}/2$  & $k_{i_2j}=-k_{i_2j}/2$; $k_{ji_2}=-k_{ji_2}/2$ \\
		end  & end & end \\
		\hline
	\end{tabular}
\end{table}
\noindent where $i_1=21$ and $i_2=20$ based on (\ref{arrange dofs}). Executing all the above pieces of codes is equivalent to performing the following matrix multiplication.

\begin{equation}\label{matrixmodification}
\begin{bmatrix}
{\bf D}^v & {} &{} \\
{}  & {\bf D}^v & {} \\
{}  & {} & {\bf D}^p \\
\end{bmatrix}
^{\rm T}
{\bf K}_e
\begin{bmatrix}
{\bf D}^v & {} &{} \\
{}  & {\bf D}^v & {} \\
{}  & {} & {\bf D}^p \\
\end{bmatrix}
.
\end{equation}

The modification of the mass matrix is similar but easier if a lumped mass is adopted, though it is unnecessary to present details here. Once the element matrix is modified, it can then be assembled directly to the global matrix and therefore implement the constraint of the hanging nodes, because the hanging node shares the same equation number with its related node in the neigbouring element.

\end{document}